\documentclass[letterpaper,twocolumn,english,aps, superscriptaddress]{revtex4-1}
\pdfoutput=1
\usepackage[T1]{fontenc}
\usepackage{booktabs}
\usepackage{units}
\usepackage{amsmath}
\usepackage{graphicx}
\usepackage{esint}

\makeatletter

\pdfpageheight\paperheight
\pdfpagewidth\paperwidth

\providecommand{\tabularnewline}{\\}
\newcommand{\lyxdot}{.}

\usepackage{enumitem}
\usepackage{lmodern}
\usepackage{color}
\usepackage{graphicx}
\usepackage {url}
\usepackage{textcomp}

\usepackage{xr}

\makeatletter\providecommand\href@noop{\@firstoftwo}\makeatother

\newcommand{\textapprox}{\raisebox{0.5ex}{\texttildelow}}

\makeatother

\usepackage{babel}

\begin{document}

\title{\textsf{Absence of selection for quantum coherence in the Fenna-Matthews-Olson
complex: a combined evolutionary and excitonic study}}

\author{St\'ephanie Valleau }

\affiliation{Department of Chemistry and Chemical biology, Harvard University,
Cambridge, MA, 02138, USA.}
\email{svalleau@fas.harvard.edu}

\author{Romain A. Studer}

\affiliation{European Bioinformatics Institute (EMBL-EBI), Wellcome Genome Campus,
Hinxton, Cambridge, CB10 1SD, UK.}

\author{Florian H\"ase}

\affiliation{Department of Chemistry and Chemical biology, Harvard University,
Cambridge, MA, 02138, USA.}

\author{Christoph Kreisbeck}

\affiliation{Department of Chemistry and Chemical biology, Harvard University,
Cambridge, MA, 02138, USA.}

\author{Rafael G. Saer}

\affiliation{Departments of Biology and Chemistry, Washington University in Saint
Louis, One Brookings Drive, St. Louis, Missouri 63130 United States}

\author{Robert E. Blankenship}

\affiliation{Departments of Biology and Chemistry, Washington University in Saint
Louis, One Brookings Drive, St. Louis, Missouri 63130 United States}

\author{Eugene I. Shakhnovich}

\affiliation{Department of Chemistry and Chemical biology, Harvard University,
Cambridge, MA, 02138, USA.}

\author{Al\'an Aspuru-Guzik}

\affiliation{Department of Chemistry and Chemical biology, Harvard University,
Cambridge, MA, 02138, USA.}
\altaffiliation{Canadian Institute for Advanced Research, Bioinspired Solar Energy Program, Toronto, ON M5G 1Z8, Canada.}

\email{alan@aspuru.com}

\begin{abstract}
We present a study on the evolution of the Fenna-Matthews-Olson bacterial
photosynthetic pigment-protein complex. This protein complex functions
as an antenna. It transports absorbed photons \textemdash excitons
\textemdash to a reaction center where photosynthetic reactions initiate.
The efficiency of exciton transport is therefore fundamental for the
photosynthetic bacterium's survival. We have reconstructed an ancestor
of the complex to establish whether coherence in the exciton transport
was selected for or optimized over time. We have also investigated
on the role of optimizing free energy variation upon folding in evolution.
We studied whether mutations which connect the ancestor to current
day species were stabilizing or destabilizing from a thermodynamic
view point. From this study, we established that most of these mutations
were thermodynamically neutral. Furthermore, we did not see a large
change in exciton transport efficiency or coherence and thus our results
predict that exciton coherence was not specifically selected for.
\end{abstract}

\maketitle

The Fenna-Matthews-Olson complex is a light-harvesting protein complex
found in green sulfur bacteria. These bacteria first appeared about
1.6 bilion years ago, in the Proterozoic era \citep{Brocks2005}.
Green sulfur bacteria survive through anoxygenic photosynthesis; they
use sulfide and other reduced sulfur compounds or hydrogen as photosynthetic
electron donors \citep{blankenshipbook}.

In the last decade, much research has been dedicated to understanding
the excitation energy transfer in the Fenna-Matthews-Olson (FMO) protein
complex \citep{Adolphs2006,Mohseni2008,Rebentrost2009b,Ishizaki2009c,Rebentrost2009a,Panitchayangkoon2010,Rebentrost2011a,Ritschel2011,Shim2012,Valleau2012b,Dostal2016}.
Once absorbed, photons become molecular excitations, or excitons.
These excitons are then transported through the complex due to their
interactions with neighboring excitations, the protein scaffold and
the environment. A renewed interest in this complex arose when low
temperature 2D-spectroscopy experiments suggested the presence of
quantum coherence effects \citep{Engel2007a,Scholes2016} in the exciton
dynamics. Quantum coherent effects can be thought of as concurrent
beats between electronic excitations which do not occur classically.
Following these experiments, much theoretical work was carried out
\citep{Fassioli2010,Ishizaki2009c,Kreisbeck2012a,Olaya-Castro2008,Olbrich2011a}
with the goal of understanding whether quantum effects were present
and if so, how they contributed to the exciton transport. 
\begin{figure}[h]
\centering{}\includegraphics[width=0.8\columnwidth]{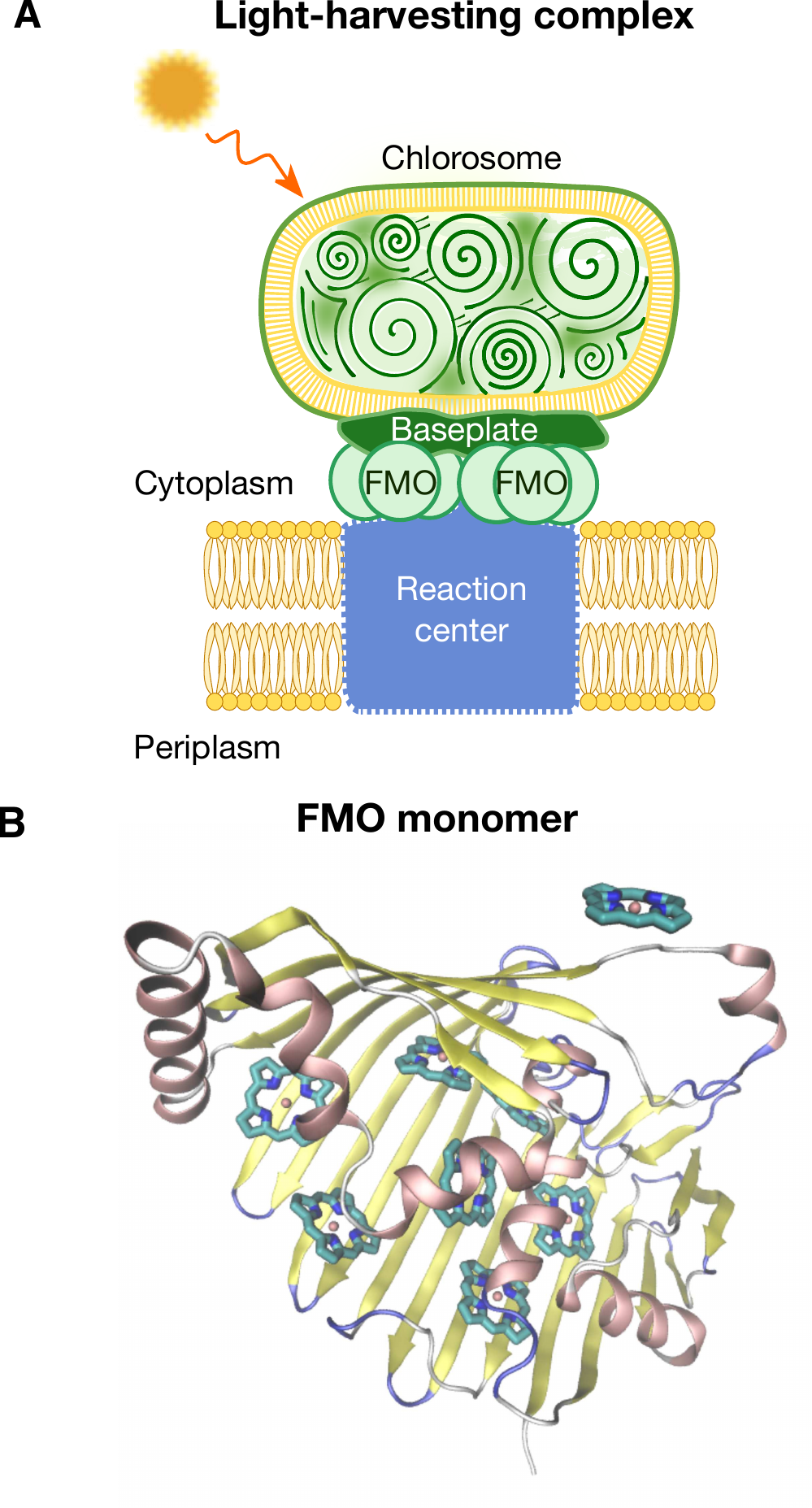}\caption{In panel \textbf{A}, an image describes the location of the FMO complex
in the light-harvesting complex of green sulfur bacteria. The photons
are absorbed by the chlorosome and transported as excitations to the
FMO complex and ultimately to the reaction center. In panel\textbf{
B} one can see crystal structure of FMO complex monomer for\emph{
Chlorobaculum tepidum} (PDB: 3ENI). The protein scaffold is colored
in pink for alpha helices, in yellow for beta sheets, and in blue
for loops. The 8 bacteriochlorophyl-a (BChl-a) pigments are colored
in cyan (their side chains are not shown for simplicity). \label{fig:iniziale}}
\end{figure}
 Most of the studies have relied on the existence of an X-ray crystal
structure of the FMO complex of either \emph{Chlorobaculum tepidum}
or \emph{Prosthecochloris aestuarii}.

FMO is found in all of the anaerobic Chlorobi phyla and recently it
has also been found in aerobic \emph{Chloracidobacterium thermophilum}
of the Acidobacteria phyla \citep{Wen2011}. In these organisms, the
FMO complex forms a homo trimer (see Fig.\,\ref{fig:iniziale}).
Each monomer consists of 7/8 bacteriochlorophyll-a (BChl-a) pigment
molecules enclosed in a protein scaffold. The BChl pigment molecules
interact with the protein scaffold through coordination and hydrogen
bonds. The complex is thought to act as an excitonic wire, funneling
the excitation from the baseplate to the reaction center where charge
separation occurs. This charge separation enables reactions which
are fundamental to the organisms' survival.
The biological origin of the complex remains a mystery. Olson et al.
(Ref. \citep{Olson2004a}) speculated that the FMO complex might have
come from an ancient reaction center. They looked for homology between
the FMO protein and PscA, the reaction center protein of green sulfur
bacteria. They found a signature sequence common to both, however
the sequences only had a 13\% identity score. More recently, the role
of some specific site mutations \citep{Saer2017288} and of cysteines
in the FMO protein has been investigated as well \citep{Saer20161455}.
In Ref. \citep{Saer20161455} it was found that cysteines appear to
be fundamental for the photoprotection mechanism of the protein complex. 

Regarding exciton transport in this system, there have only been few
efforts to understand whether and how protein evolution might have
influenced it. For instance, in Ref.\,\citep{Muh2007}, the authors
computed the BChl transition energies and looked at the effect of
including the point charges coming from the protein. They also looked
at how changes in the polar groups of amino acids, such as the amino
or hydroxyl groups, influence BChl transition energies. They found
that the charges coming from the alpha helices seem to influence the
energies strongly. On the other hand, no large change in the energies
was observed when they modified the charge distribution of single
amino acids as a means to emulate single point mutations. Experimentally,
the Blankenship group \citep{Tronrud2009a} looked at comparing the
optical properties of FMO coming from three different species to understand
the effect of the protein scaffold. They observed spectral differences,
and these were assigned to the way the BChl molecules bind to the
protein scaffold in each species. This suggests that one may probe
exciton dynamics and the biological role of coherence through point
mutations. In this sense, we are most interested in mutations occurring
in the vicinity of BChl binding sites as the effect of their charge
is expected to have a stronger influence on the BChl energies.

For the above reasons, we have carried out a theoretical study of
the protein complex from an evolutionary perspective. We aimed to
identify whether quantum coherence in exciton transport and thermodynamic
stability (free energy change upon folding) guided the evolution of
the protein complex. Hence, we carried out calculations which trace
evolution of exciton transport as well as evolution of thermodynamic
stability. For the thermodynamics, we studied the stability of the
current day protein complex to mutations by computing and comparing
the folding free energy change of the mutant to that of the non mutated
protein complex. Subsequently, we constructed an ancestor for the
FMO complex and computed the folding free energies due to point mutations
along the phylogenetic tree. Thus, with these two results, we were
able to compare the stability of \emph{C. tepidum} to the stability
of it's ancestors and see how it changed along the tree. To investigate
on role of exciton transport, we computed and compared exciton transport
in the ancestor with that of \emph{C. tepidum}. Because two-dimensional
photon echo (2D-echo) spectroscopy is the experimental tool employed
to reveal coherent phenomena, we computed this spectrum to determine
the role of coherence in the evolution from ancestor to \emph{C. tepidum}.
Finally, from the reconstructed phylogenetic tree, we identified sites
under positive selection and determined whether positively selected
amino acids were close to BChl molecules and could influence exciton
transport. 

\section*{\textsf{\large{}Results}}

\subsection*{\textsf{Robustness of current day FMO proteins to mutations}}

\begin{figure}
\centering{}\textbf{\textit{\includegraphics[width=1\columnwidth]{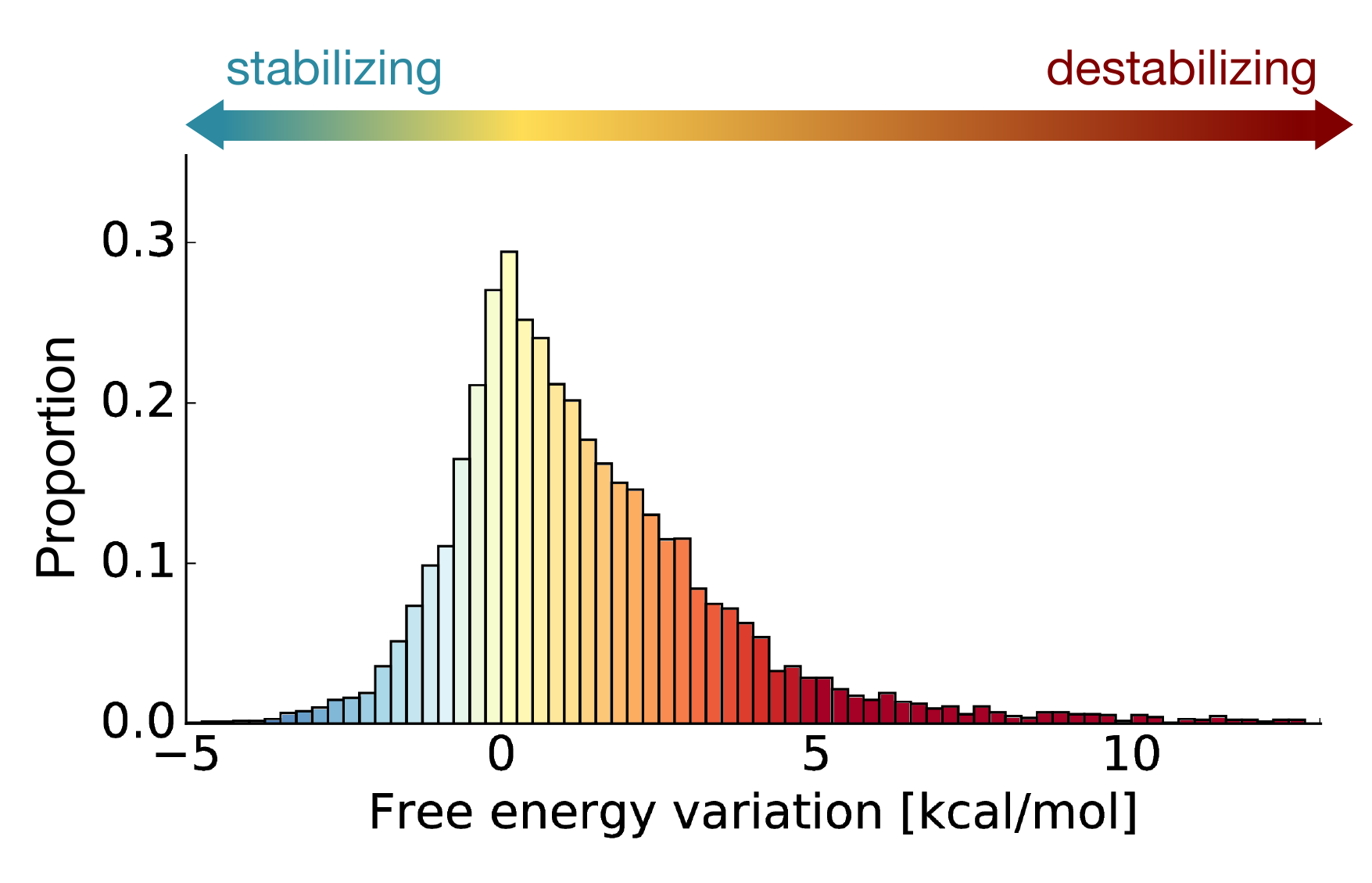}}}\caption{Histogram of free energy variation upon folding, $\Delta\Delta G=\Delta G_{mutant}^{fold}-\Delta G_{wild}^{fold}$,
for all possible single-point mutations in the FMO complex trimer
of \emph{P. aestuarii} (PDB: 3EOJ). The coloring is shaded to be more
red for destabilizing mutations (positive values) and blue for stabilizing
mutations (negative values). The bins in the histogram are normalized
by the area under the histogram. \label{fig:hist_dG}}
\end{figure}
We computed the free energy change upon folding, $\Delta\Delta G=\Delta G_{mutant}^{fold}-\Delta G_{wild}^{fold}$,
as a measure of robustness of the FMO complex to mutations. Calculations
were carried out using FoldX \citep{Guerois2002} for the trimers
of \emph{C. tepidum} and \emph{P. aestuarii} for all possible single
point mutations (see Methods for details). In our model, each single
point mutation occurs at the same time in all monomers of the trimer.
We found that most single point mutations would be destabilizing \textapprox{}~
60\% (see Fig.~\ref{fig:hist_dG}). The overall landscape of mutations
is the same for \emph{C. tepidum} and\emph{ P. aestuarii} (see Fig.~\ref{fig:hist_dG-1}
in the Supporting Information). This is qualitatively similar to the
general statistics of the effects of mutations on stability (see Ref.
\citep{Zeldovich09102007,Tokuriki20071318}).

We also looked at how the free energy change varies depending on the
location of the mutation in the protein complex monomers of \emph{P.
aestuarii }and \emph{C. tepidum} and found that mutations at the interface
of monomers are among the most destabilizing (see Fig. \ref{fig:DDG_landscape-1}).These
observations rely on the assumption that each single point mutation
(in each monomer) does not impede the assembling of a trimeric structure.

\subsection*{\textsf{Ancestor reconstruction, selection of mutations and variation
of the free energy landscape}}

\begin{figure*}
\centering{}\includegraphics[width=1.9\columnwidth]{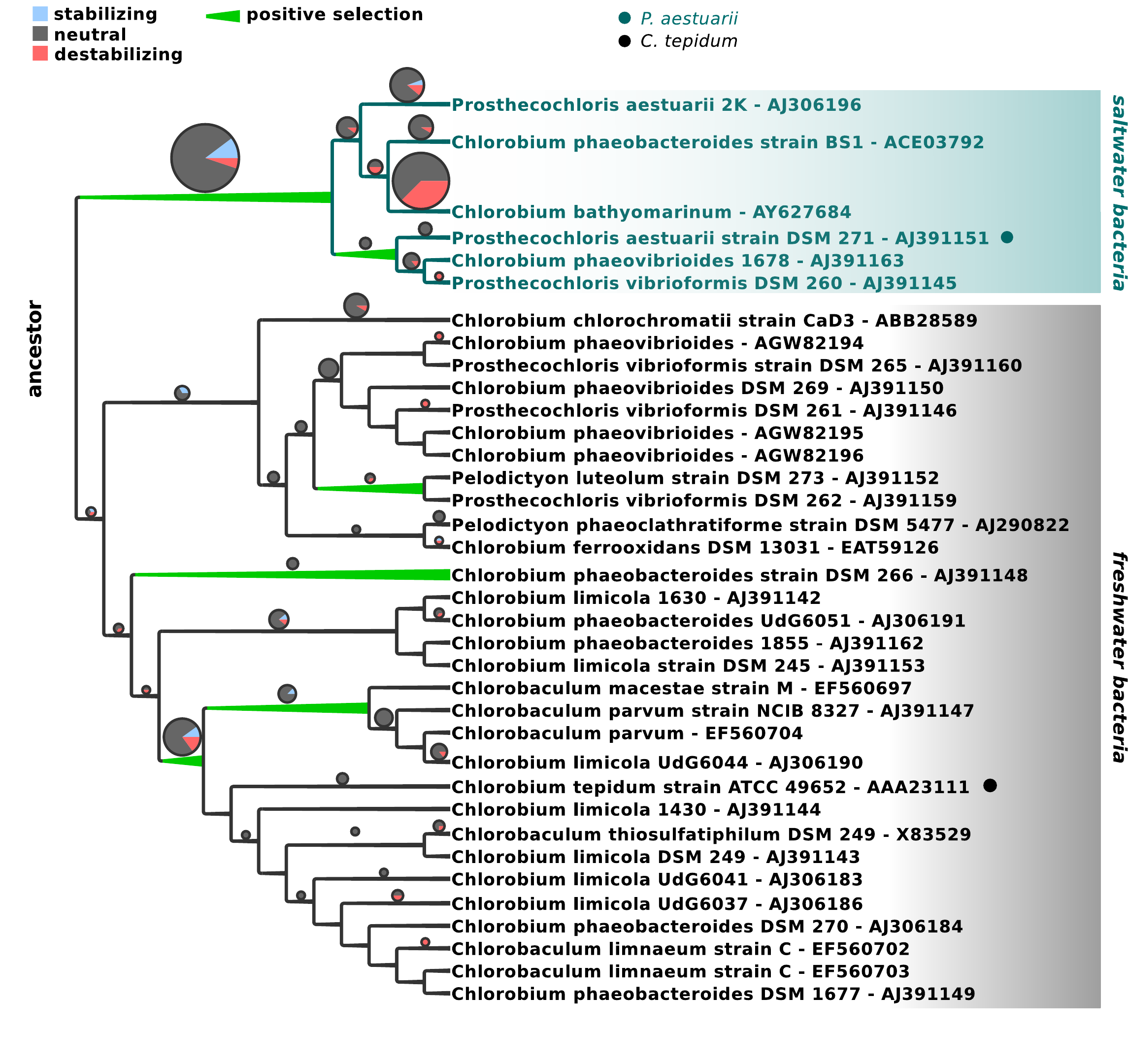}\caption{Rooted cladogram for \emph{fmoA} computed using MAFFT and TranslatorX
\citep{Abascal2010} for the alignment and PHYML \citep{Guindon01052010}.
Above each branch we show a pie-chart label with a diameter proportional
to the number of single point mutations and color based on the associated
free energy change $\Delta\Delta G$. Red represents destabilizing
mutations, blue represents stabilizing mutations, and grey represents
neutral mutations. \label{fig:Phylogenetic_tree}}
\end{figure*}
\begin{figure*}
\begin{centering}
\includegraphics[width=2\columnwidth]{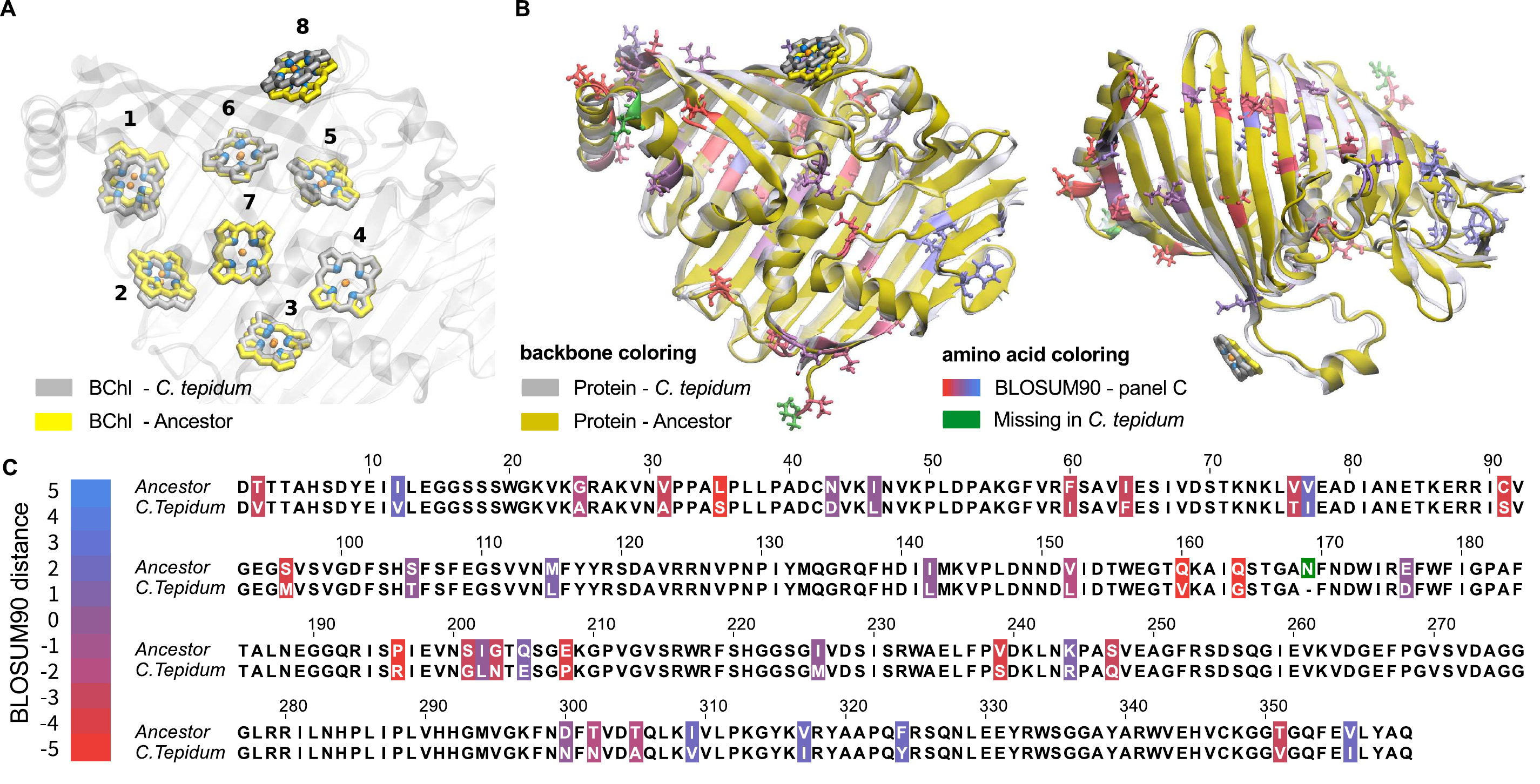}
\par\end{centering}
\centering{}\caption{In panel \textbf{A}, we show the structure of the BChl molecules in
a monomer of the ancestral structure (yellow) overlaid with the BChl
molecules in the current day structure of \emph{C. tepidum }(light
grey). The images were generated from snapshots of the molecular dynamics
simulations at 300K. In panel \textbf{B}, for the same snapshots we
compare the backbone of the ancestral FMO structure (yellow) to that
of the current day \emph{C. tepidum} structure (light grey backbone).
Some amino acids are colored based on the BLOSUM90 matrix, which quantifies
how different they are (positive numbers indicate easy substitution
while negative numbers indicate difficult substitution). These amino
acids are also highlighted in the legend in panel \textbf{C}. The
structures are snapshots from the molecular dynamics simulations at
300K. The amino acids which differ in the two structures are colored
based on the BLOSUM90 matrix which quantifies how likely it is to
substitute one amino acid with the other. The amino acid mutants in
the ancestral structure are also shown using the bond and atom representation.\label{fig:Ancestor_vs_Ctepidum}}
\end{figure*}
A phylogenetic tree was generated using the multiple sequence alignment
built from the \emph{fmoA} amino acid sequences (see details in Methods).
The amino acid sequences were obtained by translating the \emph{fmoA}
monomer gene sequences. In Fig. \ref{fig:Phylogenetic_tree} we show
the cladogram tree (phylogenetic tree is also shown in Fig. \ref{fig:Rooted-phylogenetic-tree-SI}).
The tree divides in two main clades: saltwater (e.g. \emph{P. aestuarii})
and freshwater (e.g. \emph{C. tepidum}) bacteria. 

At each node of the tree, the sequence and structure of the protein
scaffold were reconstructed. The ancestral amino acid sequences were
obtained using Bayesian inference and the scaffold structures were
computed using homology modeling (see Methods). The pie-charts above
each branch indicate, based on the diameter, how many mutations occurred
between the two branch nodes, and based on the color, whether they
were stabilizing (blue), neutral (grey) or destabilizing (red). Most
mutations were neutral in terms of modifying the free energy change
upon folding.

Given the inferred ancestor amino acid sequence at the root we constructed
a guess for the full ancestral structure (protein scaffold and BChl
molecules) through a combination of homology modeling and molecular
dynamics simulations (see Methods). An image of the structure of the
putative ancestor compared to that of \emph{C. tepidum} for one monomer
is shown in Fig. \ref{fig:Ancestor_vs_Ctepidum}. The images in each
panel are constructed from a snapshot of the molecular dynamics simulations
at 300K after equilibration. In panel B, we show two orientations
of one monomer of the protein complex. No large difference in the
backbone of the protein is observed. This is due to the fact that
the protein scaffold was constructed with homology modelling. The
differences come from the location of side chains and the change in
charge of the mutated amino acids. In panel B, mutations are highlighted
using a color scheme based on the BLOSUM90 distance. We see a few
red mutations (large negative BLOSUM90 distance) on the beta sheet
which could be a problem for the formation of the trimer. Another
effect of the mutations is the change in orientation of BChl molecules
(see Panel A). This change in orientation of the BChls influences
the coupling amongst their first excited states (see Table \ref{tab:Site-couplings-in-1}).
In particular, the coupling between the first excited states of BChl
3 and 4 is weaker in the ancestor than in \emph{C. tepidum} and, on
the other hand, the coupling between the first excited states of BChl
4 and 5 is stronger in the ancestor than it is in \emph{C. tepidum}.
Finally, as can be seen in Tab. \ref{tab:Average-distances-and},
the average distance of the amino acids from the BChl molecules changes
between the ancestor and \emph{C. tepidum} and the differences are
on the order of 0.1-0.4$\mathrm{\AA}$. 

\begin{figure}
\centering{}\includegraphics[width=0.95\columnwidth]{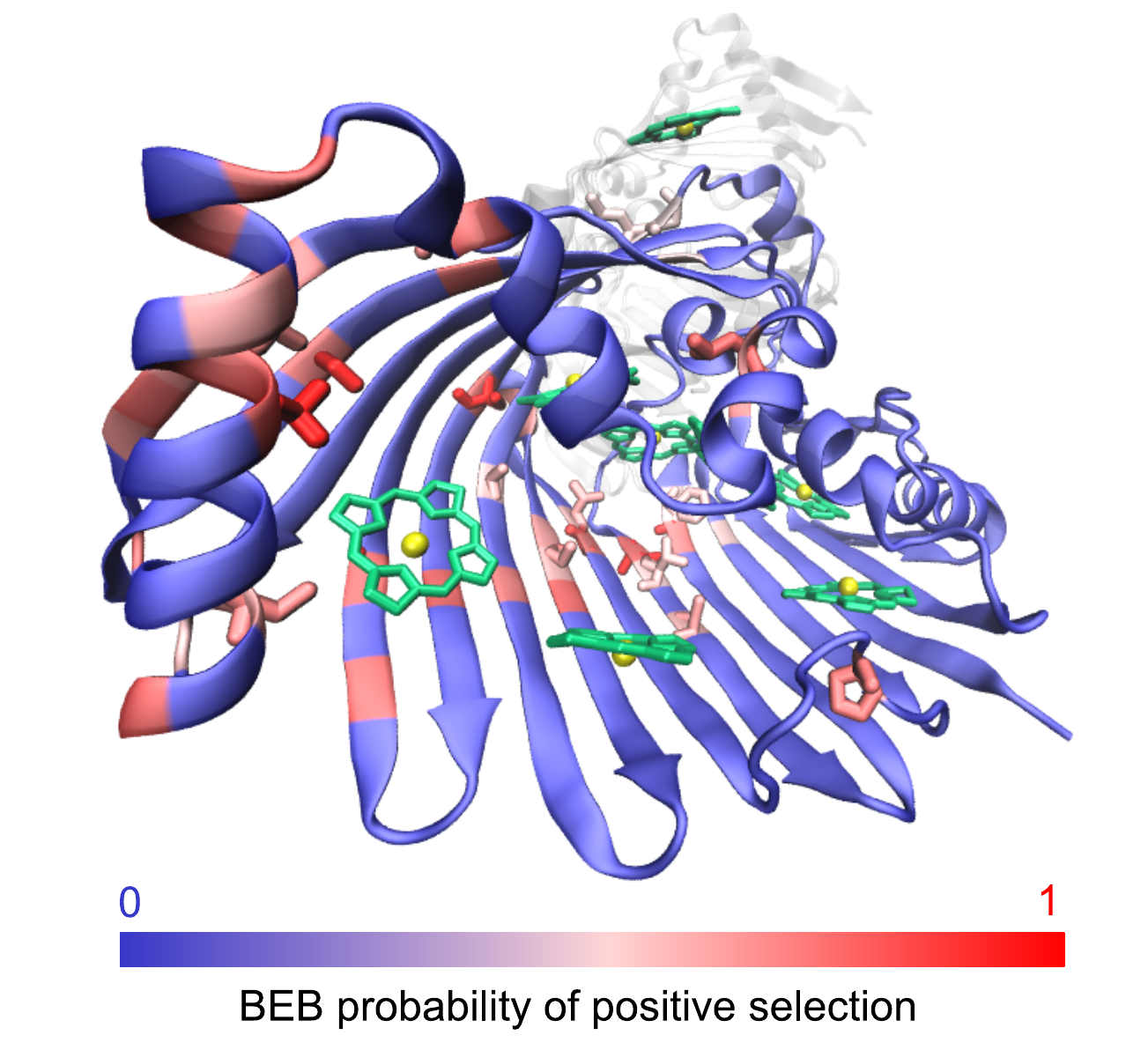}\caption{Positively selected sites shown in the structure of \emph{C. tepidum}
(PDB: 3ENI). The color gradient corresponds to the strength of positive
selection (darkest red - strongest selection, blue - no positive selection).
We only plot the side chains of positively selected residues which
are within $6\:\protect\AA$ of the magnesium in the BChl molecules.\label{fig:Positively-selected-sites}}
\end{figure}

Given the phylogenetic tree, we also identified branches under positive
selection (bright green) by using the branch-site model (details in
Methods). In Fig. \ref{fig:Positively-selected-sites}, we show mutations
which were positively selected for along these branches in the structure
of \emph{C. tepidum}. 
 Most positively selected mutations are distant from the BChl residues,
and therefore we currently do not believe they had a strong role in
modifying exciton transport. Given the high computational cost associated
with constructing the full structure (with BChl molecules) at each
node, we did not investigate the role of each positively selected
mutation using atomistic \emph{ab initio} methods. However, we computed
the free-energy changes associated with the mutations. 

\begin{figure}
\centering{}\includegraphics[width=1\columnwidth]{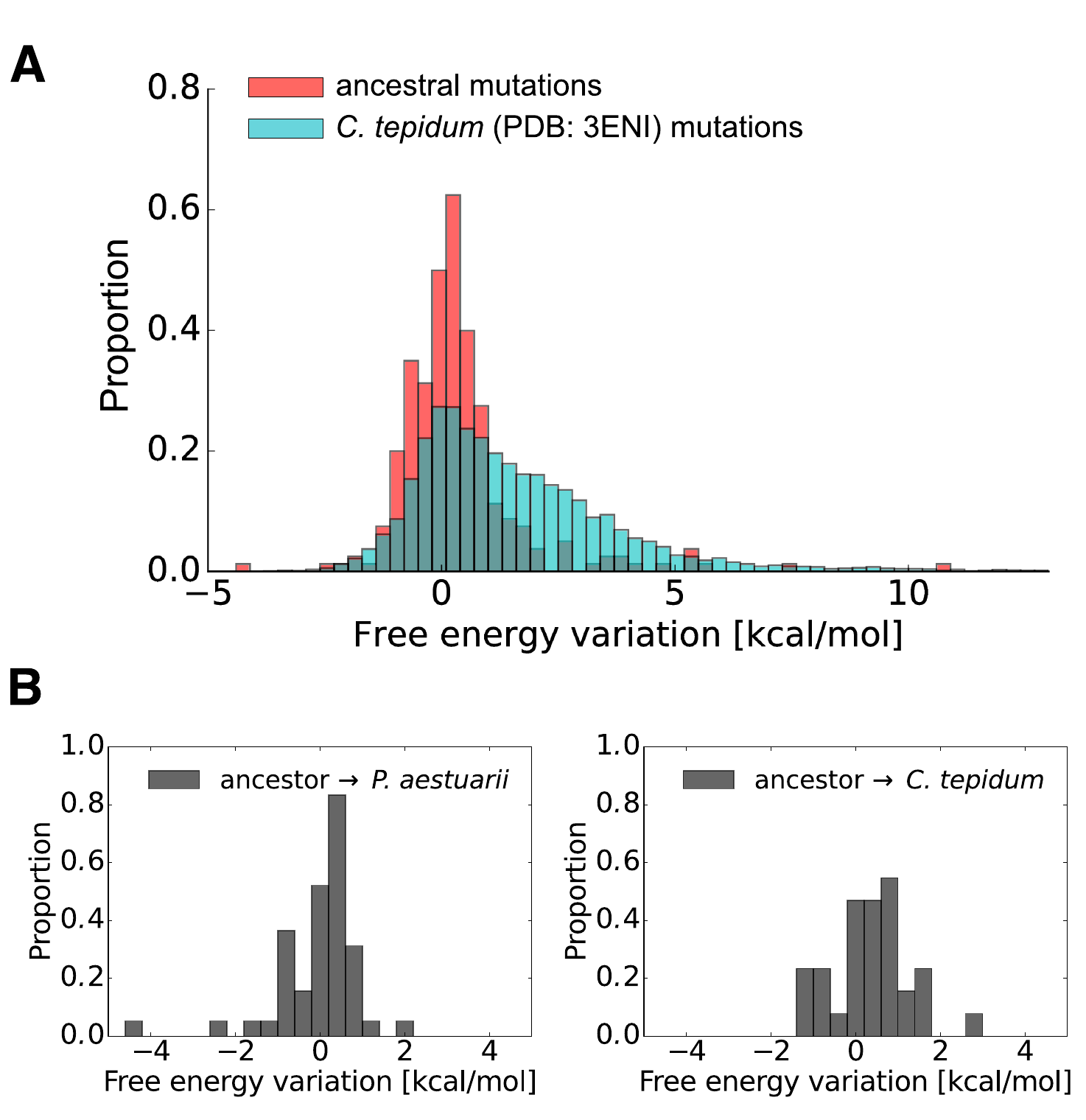}\caption{Panel \textbf{A}: In blue we show a histogram of the folding free
energy changes for all possible single point mutations of the current
day \emph{C. tepidum }FMO complex structure. In red, we show a histogram
of the folding free energy changes for all the single point mutations
which are necessary to connect the ancestral FMO structure to the
current day structures (i.e. all single point mutations along all
branches). Panel \textbf{B}: Right-hand side: histogram of the folding
free energy changes for the mutations which connect \emph{P. aestuarii}
to the ancestor. Left-hand side: histogram of the folding free energy
changes for the mutations which connect \emph{C. tepidum} to the ancestor.
Note that, for all histograms, each bin contains the number of points
with energy in that interval normalized by the area under the entire
histogram. \label{fig:Free-energy-landscape-ancestor-1}}
\end{figure}

In Fig. \ref{fig:Free-energy-landscape-ancestor-1}, panel A, we show
the histogram of free energy changes which connects the ancestor to
current day species\emph{ }in red and, in cyan, the histogram of folding
free energy variation for all possible single-point mutations in the
FMO complex trimer of \emph{C. tepidum. }We found that 65\% of mutations
are destabilizing for \emph{C. tepidum.} On the other hand only 32\%
of mutations of the ancestral FMO complex along the phylogenetic tree
were destabilizing while 21\% were stabilizing and 47\% were neutral.
This could indicate that mutations which favored the folding free
energy change variation during evolution led to more stable current
day protein structures. However, we cannot claim this in general as
the trend changes for different current day species. For instance,
as can be seen by comparing the two figures in panel B of Fig. \ref{fig:Free-energy-landscape-ancestor-1},
the distribution of folding free energy changes for mutations which
connect the ancestor to \emph{P. aestuarii} is significantly different
from that for mutations which connect the ancestor to \emph{C. tepidum}.
This difference was confirmed by the Kolmogorov-Smirnov test. Furthermore,
we computed that, for the mutations which connect the ancestor to
\emph{P. aestuarii}, 25 \% are stabilizing and 19\% are destabilizing.
However, in the case of mutations which lead to \emph{C. tepidum},
47\% are destabilizing while 22\% are stabilizing. This indicates
that \emph{C. tepidum} is less stable than the ancestor while \emph{P.
aestuarii} is more stable than the ancestor.

\subsection*{\textsf{Exciton transport and coherence}}

\subsubsection*{\textsf{Absorption, circular dichroism and linear dichroism}}

\begin{figure}
\centering{}\includegraphics[width=0.6\columnwidth]{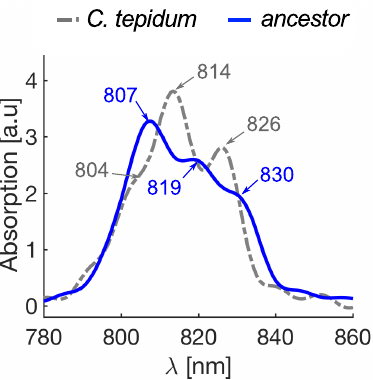}\caption{Comparison of simulated absorption spectra of the ancestral FMO complex
(blue line) to that of current day \emph{C. tepidum} at 300K (grey
dashed line) in arbitrary units. Spectra were computed using the Qy
transitions obtained by using QM/MM. See Methods for more details.
\label{fig:Comparison-of-absorption}}
\end{figure}

In Fig. \ref{fig:Comparison-of-absorption}
 we compare the absorption spectrum of the ancestor to that of \emph{C.
tepidum}. The ancestor spectrum shows three main absorption peaks
at 807, 819 and 830$\,\text{nm}$ while \emph{C. tepidum} shows strong
absorption at 804, 814 and 826$\,\text{nm}$. These trends and the
corresponding circular dichroism spectra (see Supporting Information)
of the ancestor and of \emph{C. tepidum} are very similar, respectively,
to the spectra of FMO in \emph{C. thiosulfatophilum} and \emph{C.
limicola} reported in Ref. \citep{Olson1976}. However, we do not
see strong evidence of a similarity between the ancestor spectrum
and the reaction center absorption spectra reported in that work.
The small bump at about 845 nm (see Fig. \ref{fig:Comparison-of-absorption})
is most likely due to noise, as it does not correspond to any of the
frequencies of the equivalent Fermi-Golden rule absorption spectrum.
Therefore, our current results do not confirm that the ancestor is
related to current day PscA, the reaction center protein of green
sulfur bacteria, however we cannot exclude that it may have been related
to ancestral reaction centers.

In the Supporting Information we also show the comparison of our \emph{C.
tepidum} spectra to the experimental absorption, linear dichroism
and circular dichroism spectra. 

\subsubsection*{\textsf{Coherence and 2D spectra}}

\begin{figure*}
\centering{}\includegraphics[width=2\columnwidth]{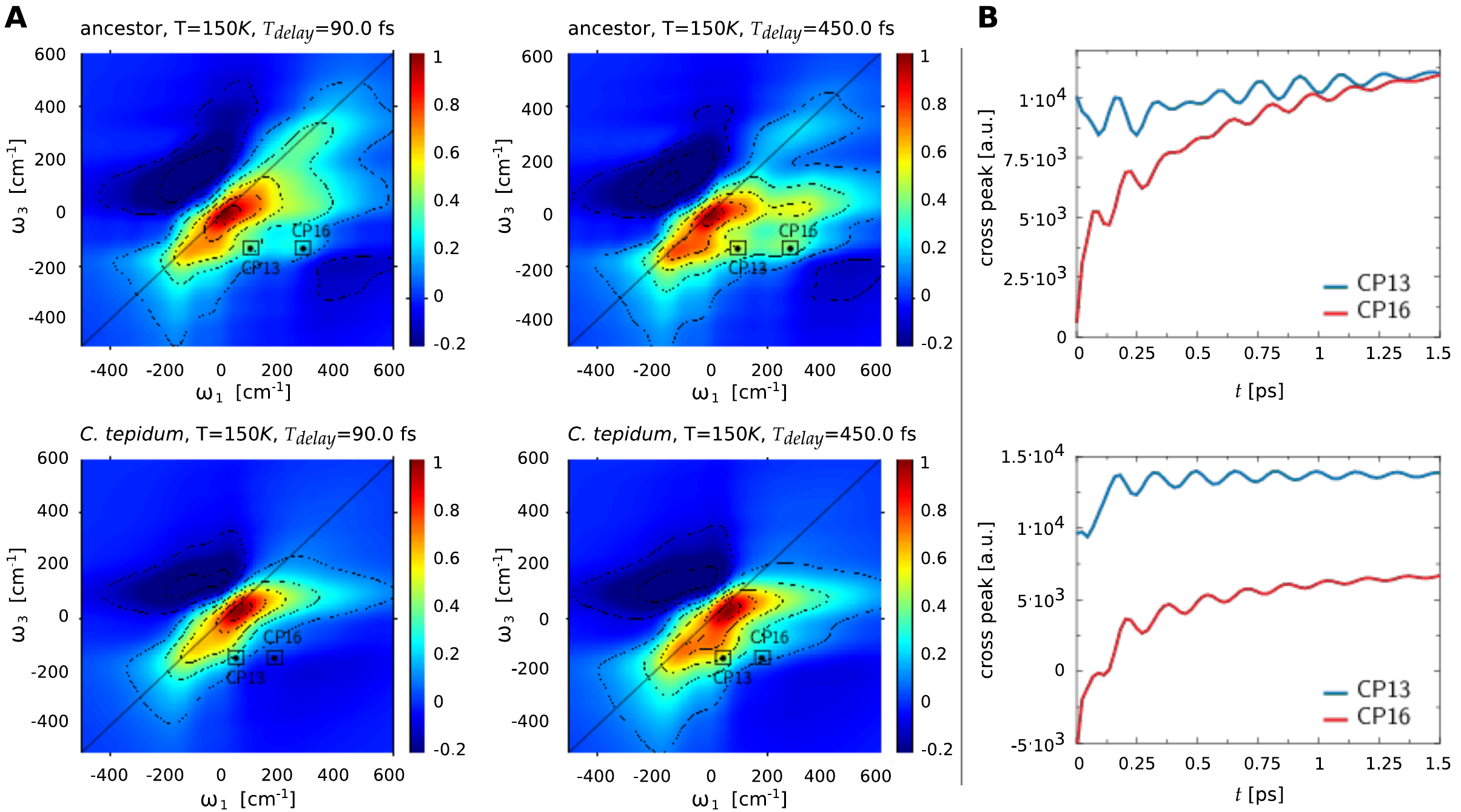}\caption{Panel \textbf{A}: Contour plots of the total computed signal for the
2D-echo spectra at different delay times. The figures in the top row
show results for the ancestor at a delay time of 90 and 400$\,$fs
and the bottom figures show the analogous plots for \emph{C. tepidum}.
Panel \textbf{B}: The amplitude of oscillation of two specific cross
peaks. The top image is for the ancestor and the bottom image is for
\emph{C. tepidum}. \label{fig:2d-echo}}
\end{figure*}

Using the hierarchy equation of motion (HEOM) non-Markovian master
equation method \citep{Tanimura1989,Tanimura2006,Ishizaki2009a},
we computed the 2D-echo spectra of the reconstructed ancestral FMO
complex and of \emph{C. tepidum} at 150 K. To this end we employed
the \textit{QMaster} software package \citep{Kreisbeck2014} (for
more details see Methods). The spectral densities which were employed
are described in Methods. These were obtained by fitting the original
QM/MM spectral densities to Drude-Lorentz spectral densities with
three peaks. The spectral densities indicated a similar coupling strength
of the system to the bath for the ancestral and \emph{C. tepidum}
FMO complexes for almost all modes. The total signal of the 2D-echo
spectra, i.e. sum of the stimulated emission (SE), ground state bleaching
(GB), and excited state absorption (ESA) pathways is shown in Fig.
\ref{fig:2d-echo}. From the oscillations of the cross-peaks in panel
B) we cannot deduce an improvement in quantum coherence or a significant
difference between the ancestor and \emph{C. tepidum}. For all species,
cross-peak beatings in the 2D-echo spectra are dominated by ground
state vibrations \citep{Kreisbeck2013} superposed by minor vibronic
contributions. This suggests that optimizing exciton transport through
quantum coherence was not part of the evolution of the FMO complex.

\section*{\textsf{\large{}Conclusion}}

We have reconstructed a structure for the ancestral Fenna-Matthews-Olson
complex and found that mutations of it's amino acids over time have
influenced the free energy upon folding of the protein complex. In
particular, the current day specie of\emph{ P. aestuarii} was obtained
through stabilizing mutations while that of \emph{C. tepidum }is less
stable than the ancestor. Regarding exciton transport, we do not observe
a significant change or improvement in the efficiency of exciton transport
or quantum coherent transport. These results suggest that the complex
did not evolve to optimize for quantum coherent transport. It is also
possible that the ancestor was already sufficiently optimized for
exciton transport and therefore further evolution or optimization
was not necessary. One of the results we find is that mutations most
likely led to better binding of the FMO complex to the baseplate and
reaction center. This is confirmed by the larger instability associated
with mutating residues in those locations. This leaves an open question:
perhaps the overall exciton transport, from the chlorosome to the
reaction center, was optimized by favoring mutations in FMO which
led to a stronger binding of FMO to its neighbouring protein systems. 

For future work, we are interested in looking at the exciton transport
at each node of the phylogenetic tree and further in expressing the
ancestral sequence in current day \emph{C. tepidum} to obtain an experimental
structure and 2D-echo spectra for the ancestor. Finally it would be
interesting to look at the role of photoprotection to see whether
the mutations could have influenced it.

\section*{\textsf{\large{}Methods and computational details\label{sec:Methods}}}

\subsection*{\textsf{Stability of the}\textsf{\emph{ C. tepidum}}\textsf{ and
}\textsf{\emph{P. aestuarii }}\textsf{Fenna-Matthews-Olson complex
to mutations}}

Each of the amino acids of the trimer and monomer of the FMO complex
of \emph{P. aestuarii }\textit{\emph{and}}\emph{ C. tepidum} were
mutated to the each of the other possible 19 amino acids using FoldX
\citep{Guerois2002}. The corresponding free energy variation upon
folding $\Delta\Delta G=\Delta G_{\text{mut}}-\Delta G_{\text{wt}}$
was obtained for each mutation. In order to account for the effect
of mutations at the interface between monomers within the trimer,
the mutations were carried out simultaneously in each monomer of the
trimer and the resulting free energy variation was normalized by dividing
by three.

In FoldX, mutations are modeled as follows. The initial crystal structure
is optimized to remove any eventual steric clashes. Then, the residue
of interest is mutated and its nearest neighbors are mutated to themselves
and conformationally relaxed to remove any local clash. The nearest
neighbors are mutated to themselves so that their geometry may be
optimized toghether with that of the central residue. During this
procedure the backbone of the protein is kept fixed and all other
residues that are far from the one of interest are also kept fixed.
The stability, $\Delta G_{{\rm wt}}$, of this relaxed structure is
obtained by using an effective energy function. Subsequently, the
residue of interest is mutated to each of the other 19 amino acids
and its neighbors to themselves and the various $\Delta G_{{\rm mut}}$
are computed. 

The procedure described above was repeated 5 times for each mutation
to ensure that the minimum energy conformations of large residues
that have many rotamers were identified. The effective energy function
(``effective energy'' here refers to the Helmoltz free energy of
a system (protein + solvent) for a fixed protein conformation) in
FoldX has been optimized for amino acid sequences, thus all BChl-\emph{a}
molecules could not be included directly in this calculation. To account
for their presence, the structures given in input to FoldX were obtained
by homology modeling with Modeller \citep{Sali1993a}. In the Modeller
simulations, the BChl molecules were inserted and kept fixed as hard
spheres. In the FoldX optimization the residues that are known to
bind to BChl-\emph{a} molecules from the crystal structure analysis
of FMO complexes from current day species were also kept fixed. A
similar procedure was recently employed successfully for the case
of RubisCO where Mg atoms are present \citep{Studer2013}.

\subsection*{\textsf{Determination of positively selected sites: site and branch-site
models}}

\subsubsection*{\textsf{Phylogenetic tree\label{subsec:Phylogenetic-tree}}}

All \emph{fmoA} gene sequences from the EMBL database were gathered
and employed to generate an alignment using translatorX with MAFFT
\citep{Abascal2010}. Phylogenetic trees were constructed using PhyML
\citep{Guindon01052010} with 5000 bootstraps. The parameters and
settings used in PhyML were as following. The LG substitution model
was chosen. We selected to have four substitution rate categories.
The alpha parameter for the gamma distribution of sites was set to
be estimated by the code. Both NNIs and SPR methods were used to search
for the optimal tree topology and finally, tree topology, branch lengths
and rate parameters were chosen to be optimized by the code. A set
of three trees were constructed. In the first tree, all sequences
were included excluding FJ210646, as it was missing about 100 residues.
Therefore, including it would have introduced more error in the tree
reconstruction phase. For the second tree, we removed \emph{Chloroacidobacterium
thermophilum} (ABV27353) as it had the lowest sequence identity percent
and thus also introduced more uncertainty. In the third tree, \emph{Chloroherpeton
thalassium} (ACF13179) was removed as it was the second most distant
sequence from the rest. We computed ancestral sequences and searched
for positively selected sites on all three trees but we present results
for the third tree as we believe it has the smallest error. 

\subsubsection*{\textsf{Site and branch-site models\label{subsec:Site-and-branch-site}}}

The branches under positive selection were identified using the branch-site
model \citep{Zhang01122005} as implemented in CodeML. This model
allows for variation of $\omega=\frac{dN}{dS}$ (the ratio of synonymous
to non synonymous mutations) amongst branches and sites. For each
simulation, one branch is selected as a foreground branch and, using
CodeML, we computed the likelihood that the branch was under positive
selection. The input DNA sequence alignment was the same as the one
generated to deduce the phylogenetic tree. Given a branch under positive
selection, we determined which amino acids were under positive selection
by using the Bayes Empirical Bayes (BEB) model \citep{Yang01042005}.
We corrected the probability values p-values for False Discovery Rate
(FDR) by using the q-value \citep{qvalue}, as discussed in Ref. \citep{Anisimova01052007}.

\subsection*{\textsf{Reconstruction of the ancestral protein structure of FMO}\label{subsec:Reconstruction-of-the}}

\subsubsection*{\textsf{Sequence reconstruction}}

An ancestral sequence was reconstructed using maxium likelihood as
implemented in FastML \citep{Ashkenazy01072012} and the LG+G model
for amino acid substitution. The LG+G model was established to be
the best model under the Bayesian Information Criterion (BIC). With
FastML, we obtained the most probable sequences, together with the
posterior probabilities for each character and indel at each sequence
position for each internal node of the tree. As inputs, we used the
phylogenetic trees described in the previous section and rooted them
based on the most distant sequence. The alignments used in the input
were the same as those employed to generate the trees. For the last
tree that excludes \emph{C. thermophilum }and \emph{C. thalassium},
we used the midpoint root, as most sequences are similar, (see Fig.
\ref{fig:Phylogenetic_tree}) and the branching is almost identical
to that obtained when \emph{C. thalassium} or \emph{C. thermophilum
}are included. 

\subsubsection*{\textsf{Structure reconstruction and molecular dynamics}}

The ancestral structure was built by using homology modeling with
satisfaction of spatial restraints as implemented in Modeller \citep{Sali1993a}.
The method of modeling by satisfaction of spatial restraints works
as follows. Constraints are generated on the structure of the target
sequence using its alignment to the related protein structures as
a guide. The restraints are obtained based on the assumption that
the structures will be similar in the aligned regions. In MODELLER,
the form of the restraints was obtained from a statistical analysis
of the relationships between similar protein structures in a database
of 105 alignments that included 416 proteins of known 3D structure
\citep{Sali1993a}. These restraints are supplemented by stereochemical
restraints such as bond lengths, angles etc, which are obtained from
a force field. Once all these restraints are established, the model
structure is obtained in Cartesian space by minimizing the violations
of all restraints. 

We used the structure of \emph{P. aestuarii} (PDB: 3EOJ) as the homologue
and generated 100 possible structures with slow refinement. The ancestral
structure was chosen to be the one with the best molpdb factor. The
initally reconstructed structure only included the protein scaffold,
as there is no current parametrization for BChl's in Modeller. Thus,
BChl molecules were included subsequently using minimization as implemented
in NAMD \citep{NAMD2005}. The chromophores were initially positioned
as in the homologous structure and the structure was optimized by
minimizing the energy in two steps. In the first step, the backbone
was kept fixed, and for the second minimization the backbone was free
to move. The structure was subsequently equilibrated without constraints
using the Amber14 software package, \citep{amber14} with the Amber
ff99SB force field\citep{amberff99sb}. The BChl-a parameters employed
are reported in Refs.~\citep{Surya2015,Shim2012}. The protonation
states of all amino acids were determined with the H++ 3.0 software
\citep{Anandakrishnan2012}, under neutral pH conditions. All complexes
were solvated using TIP3P periodic water boxes, \citep{Jorgensen1983}
with a minimum distance of $\unit[15]{\AA}$ between the complex and
the box boundaries. Charges were neutralized by adding sodium ions
for the ancestor and \emph{P. aestuarii} and chloride for \emph{C.
tepidum}. Shake constraints were applied to all bonds containing hydrogen.
Minimizations were carried out for $2000$ steps for \emph{P. aestuarii}
and \emph{C. tepidum} and for $10000$ steps for the ancestor. Minimizations
were followed by $\unit[200]{ps}$ adaptation runs to impose a temperature
of $\unit[300]{K}$ and a pressure of $\unit[1]{atm}$ on the systems.
All three complexes were equilibrated for $\unit[50]{ns}$ in the
same environmental conditions. Long range electrostatic interactions
were calculated using the Particle-Mesh Ewald method \citep{Essmann1993}.
For each complex we then carried out $\unit[40]{ps}$ production runs
with a $\unit[1]{fs}$ integration step. 

\subsection*{\textsf{Correlating biological stability and evolution to exciton
dynamics\label{subsec:Exciton-1}}}

\subsubsection*{\textsf{QM/MM Hamiltonian}}

For the exciton dynamics, the FMO complex is simulated as a system
coupled to a bath (the protein environment). The system Hamiltonian
is defined as 
\begin{eqnarray}
H_{\text{ex}} & = & \sum_{i=1}^{8}\varepsilon_{i}|i\rangle\langle i|+\sum_{i\neq j}^{8}V_{ij}|i\rangle\langle j|\label{eq:system_hamiltonian-1}
\end{eqnarray}
where $\varepsilon_{i}$ is the first excited state energy of the
$i$-th BChl. The $V_{ij}$ terms correspond to the excitonic coupling
between the excited states for the $i-th$ and $j-th$ molecule. 

Site energies were computed using TDDFT in QChem \citep{Shao2006}
with the PBE0 functional \citep{PBE0} and 3-21G basis set for $10000$
frames ($\unit[40]{ps}$) of the MD production run. Couplings were
calculated using two approaches: the point-dipole approximation (PDA)
and via transition charges from the electrostatic potential (TrEsp) \citep{Madjet2006}.
TrEsp has shown to be the most accurate, although it is computationally
more demanding than using the PDA \citep{Madjet2006}. The computed
couplings are given in the Supplementary information.

\begin{figure}
\centering{}\includegraphics[width=1\columnwidth]{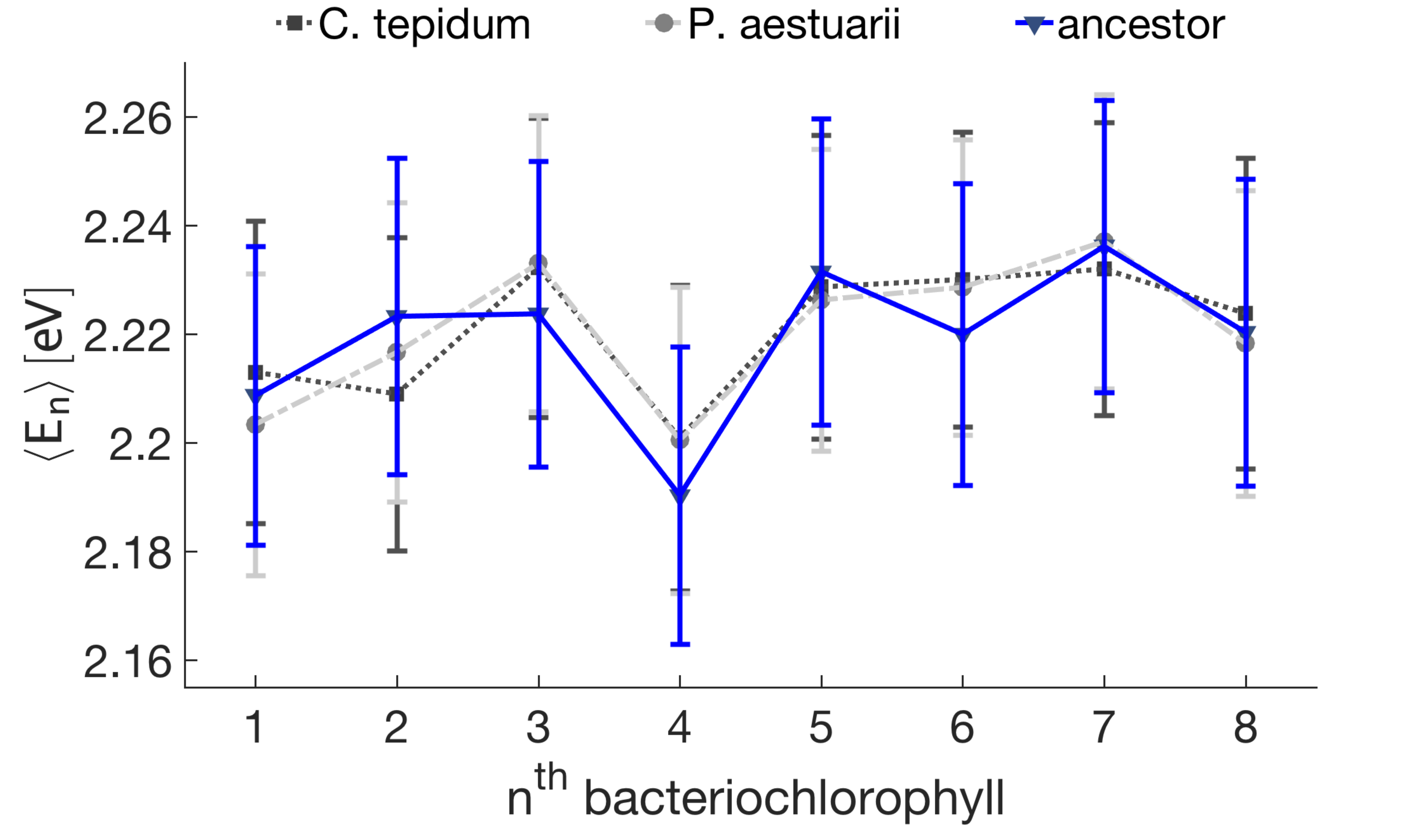}\caption{Average first excited state energies $\langle\sf{E_{n}}\rangle$for
each BChl molecule in the FMO complex of \emph{C. tepidum }(black),\emph{
P. aestuarii }(grey)\emph{ }and for the ancestral structure (blue).
Values were obtained by averaging over the production QM/MM trajectory
at 300K.\label{fig:Compare-average-energies-1}}
\end{figure}

The average first excited state energy of each BChl isshown in Fig.
\ref{fig:Compare-average-energies-1}. We notice there is no large
change in the trend of the excited state energies for the BChl's. 

\subsubsection*{\textsf{Exciton coherence and 2d spectrum}}

The 2d-echo spectra in Fig. \ref{fig:2d-echo} were calculated as
in Ref. \citep{Hein2012}. In two-dimensional electronic spectroscopy
the sample is probed by a sequence of thee laser pulses. Adjusting
the delay time $T_{delay}$ between the second and the third pulse
yields a time resolved picture of the exciton dynamics. Within the
impulsive limit (assuming $\delta$-pulses), the 2D-echo spectrum
is related to the third order response function $S^{(3)}(t_{1},T_{delay},t_{3})$.
Phase matching ensures that the components of the detected signal
go along distinct directions, and thus can be experimentally separated.
We considered the rephasing (RP) component of the signal 
\begin{align}
I_{RP}(\omega_{1},T_{delay},\omega_{3}) & =...\label{IRP}\\
 & \iintop_{0}^{\infty}\mbox{d}t_{1}\mbox{d}t_{3}e^{i\omega_{3}t_{3}-i\omega_{1}t_{1}}S_{RP}^{(3)}(t_{1},T_{delay}t_{3})\nonumber 
\end{align}
for which coherent phenomena show as unique oscillatory pattern in
the cross-peak dynamics as a function of delay time. Using the notation
of double sided Feynman diagrams \citep{Mukamel1999,Cho2009}, the
total signal is given by three separate pathways: stimulated emission
(SE), ground state bleaching (GB) and excited state absorption (ESA).
The latter involves the double exciton manifold. We numerically evaluated
$I_{RP}(\omega_{1},T_{delay},\omega_{3})$ by propagating the reduced
density matrix, describing the exciton degrees of freedom, along the
double sided Feynman diagrams. Hereby, each interaction with the laser
pulses requires an interruption of the dynamics and a multiplication
of the reduced density matrix with the dipole operator, either from
the right or from the left. The propagation was done using \textit{QMaster}
\citep{Kreisbeck2014}, a high-performance implementation of HEOM
\citep{Kreisbeck2011,KreisbeckCode}. The rotational average over
random orientations of the probed sample was included by sampling
20 laser polarization vectors aligned with the vertices of a dodecahedron
\citep{Hein2012}.

\section*{\textsf{\large{}Acknowledgements}}

S. V. would like to thank Pouria Dasmeh and Adrian Serohijos for their
interesting discussions on methods to compute the free energy changes
for mutations and phylogenetic tree reconstruction. The authors thank
Nicolas Sawaya for sharing his code for the calculation of atomic
partial charges from the transition densities. The computations in
this paper were completed on the Odyssey cluster supported by the
FAS Division of Science, Research Computing Group at Harvard University.
S. V., F. H. and A. A.-G. acknowledges support from the Center for
Excitonics, an Energy Frontier Research Center funded by the U.S.
Department of Energy, Office of Science and Office of Basic Energy
Sciences under Award Number DE-SC0001088. A.A.G. acknowledges the
generous support from the Canadian Institute for Advanced Research.
This work was supported as part of the Photosynthetic Antenna Research
Center (PARC), an Energy Frontier Research Center funded by the U.S.
Department of Energy, Office of Science, Office of Basic Energy Sciences
under Award Number DE-SC 0001035 to REB. RS was supported by the PARC
grant.{\small{}\medskip{}
}{\small \par}

{\small{}\bibliographystyle{unsrt}
\bibliography{photosynthex_ordered}
}{\small \par}

\clearpage{}

\pagebreak{}

\section*{\textsf{Supporting Information (SI)}}

\renewcommand{\thetable}{S\arabic{table}} \renewcommand{\thefigure}{S\arabic{figure}}
\setcounter{figure}{0}
\setcounter{table}{0}

\subsection*{\textsf{Phylogenetic tree}}

\begin{figure*}
\centering{}\includegraphics[width=2\columnwidth]{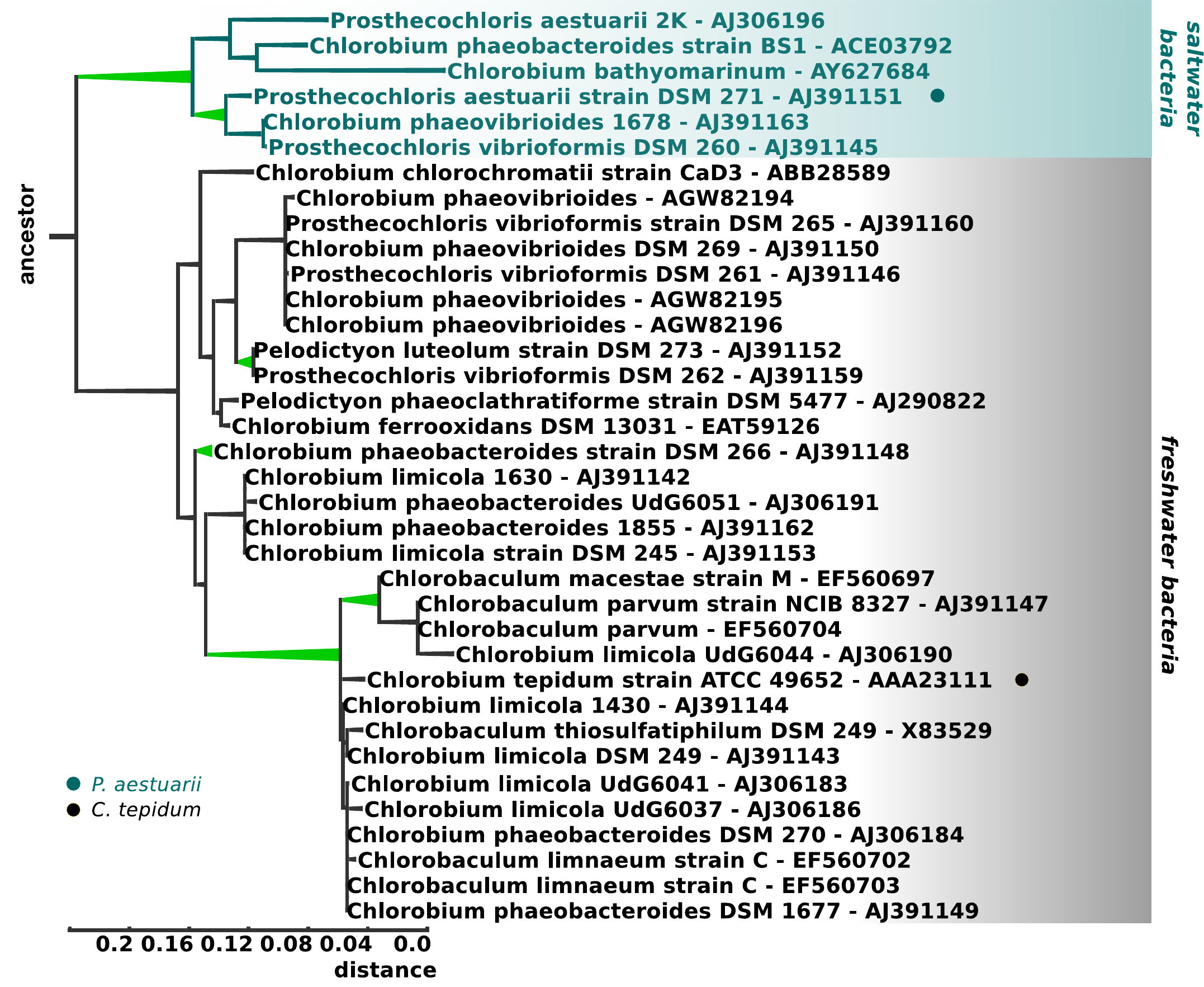}\caption{Rooted phylogenetic tree for \emph{fmoA} computed using MAFFT and
TranslatorX \citep{Abascal2010} for the alignment and PHYML \citep{Guindon01052010}.
Distances between branches correspond to difference in terms of sequence
identity. Branches that are colored in light green are branches where
positive selection occurred. \label{fig:Rooted-phylogenetic-tree-SI}}
\end{figure*}

In Fig.$\,$\ref{fig:Rooted-phylogenetic-tree-SI} we show the phylogenetic
tree of the FMO complex, based on the \emph{fmoA} gene, with branch
lengths defined in terms of sequence identity.

\subsection*{\textsf{Thermodynamic results}}

\begin{figure*}
\centering{}\textbf{\textit{\includegraphics[width=0.95\columnwidth]{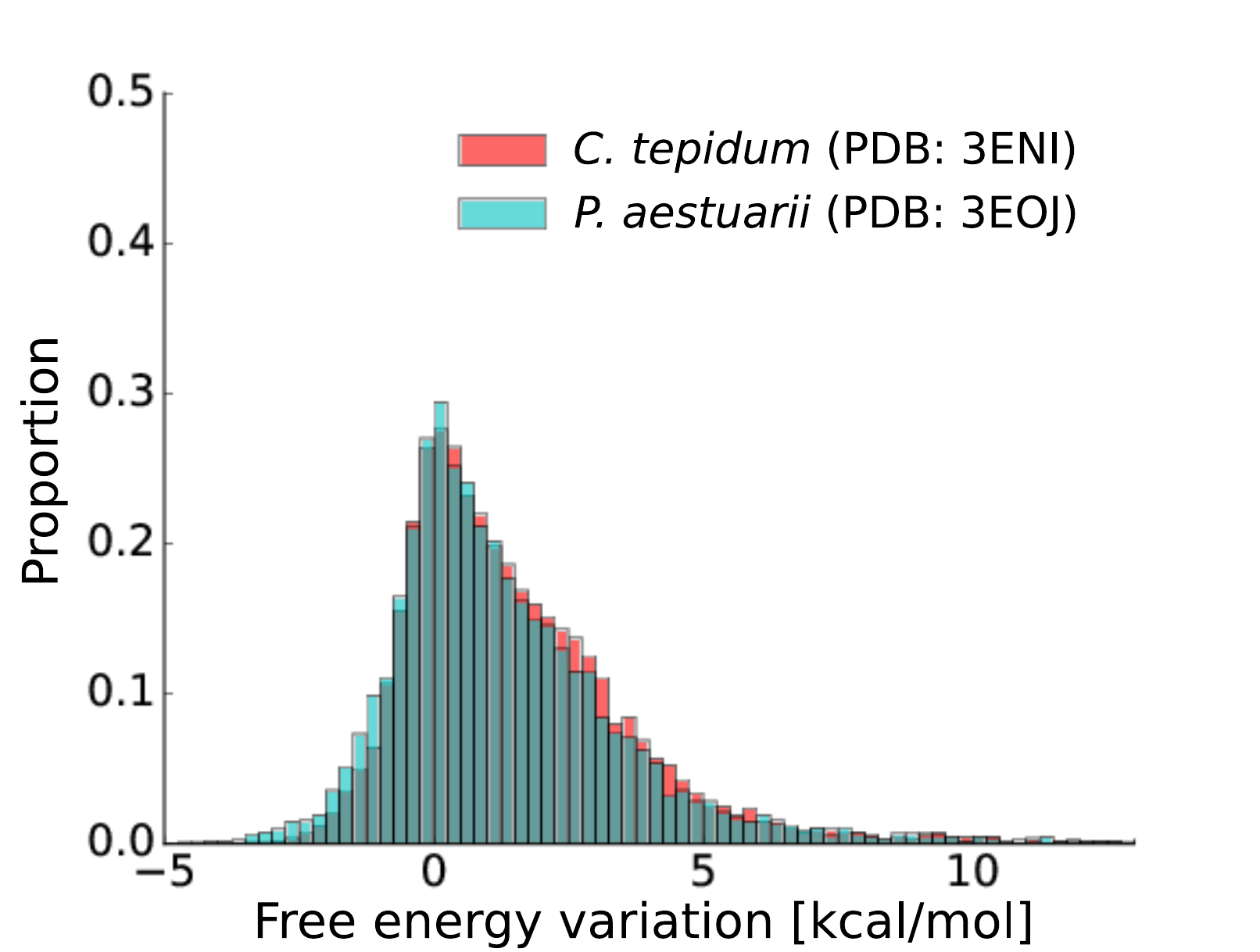}}}\caption{Overlap of free energy variation histograms for all single-point mutations
which could be applied to \emph{C. tepidum} (PDB: 3ENI, red) and \emph{P.
aestuarii} (PDB: 3EOJ, blue). Each bin contains the number of points
with energy in that interval normalized by the area under the entire
histogram.\label{fig:hist_dG-1}}
\end{figure*}

In Fig.\,\ref{fig:hist_dG-1} we compare the free energy variation
histograms for \emph{P. aestuarii} and \emph{C. tepidum}. We notice
that the distributions are not significantly different. This was also
checked by a Kolmogorov-Smirnov goodness-of-fit test. 

\begin{figure*}
\centering{}\includegraphics[width=2\columnwidth]{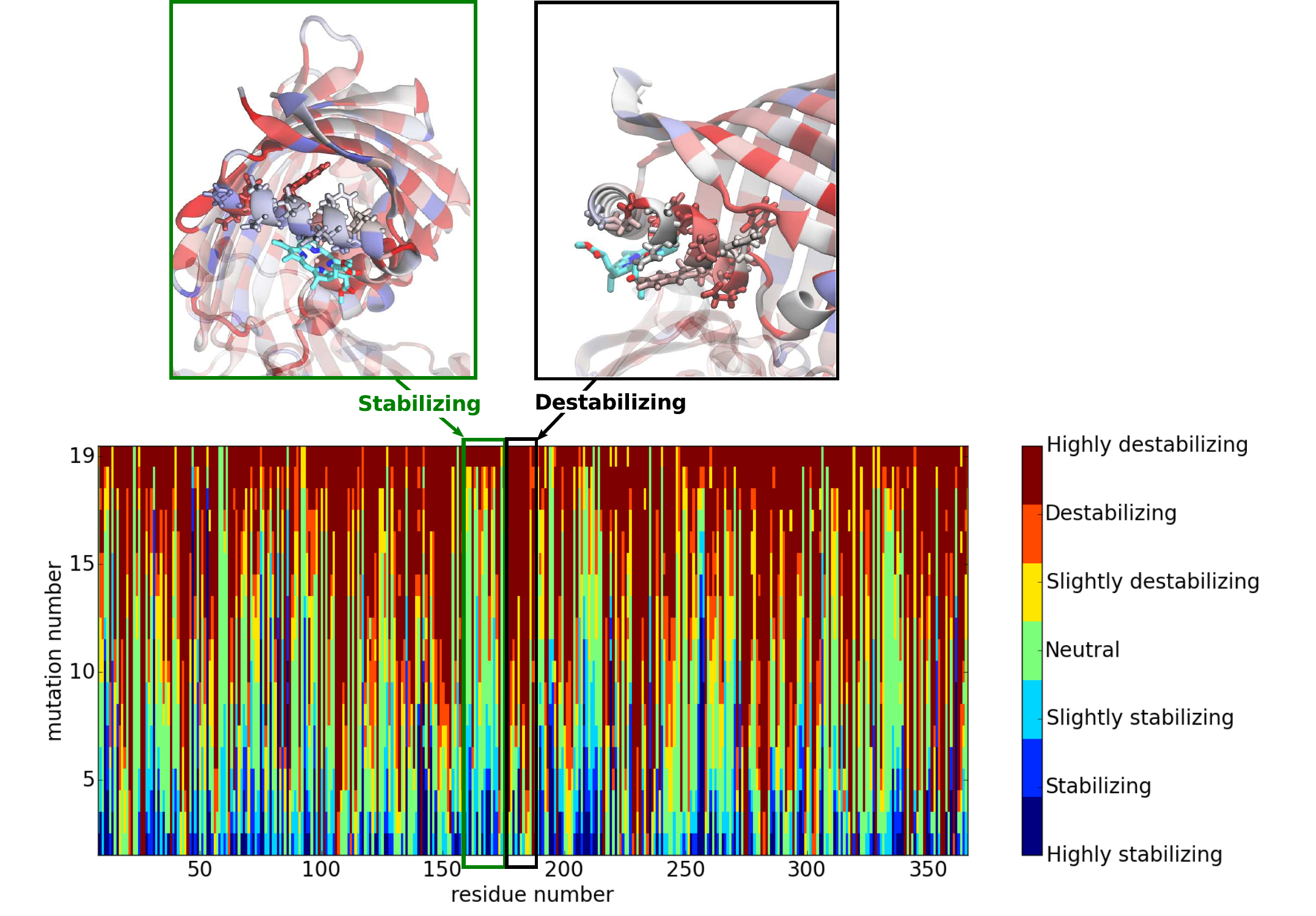}\caption{Histogram that summarizes the results of the computed $\Delta\Delta G$
for all single point mutations in \emph{P. aestuarii}. For each residue
number (x-axis) we show a color that represents the free energy change
associated with each of the 19 possible single point mutations. The
color is selected based on whether the free-energy change is stabilizing
or not. Using $\sigma=0.46\,{\rm kcal/mol}$, the free energy changes
where classified as neutral (green) for $-\sigma\leq\Delta\Delta G\leq\sigma$,
slightly destabilizing (yellow) for $\sigma<\Delta\Delta G\leq2\sigma$,
destabilizing (orange) for $2\sigma<\Delta\Delta G\leq3\sigma$ and
highly destabilizing (burgundy) for $3\sigma<\Delta\Delta G$. Similarly,
the stabilizing mutations were classified as $-2\sigma\geq\Delta\Delta G>-\sigma$
for slightly stabilizing (cyan), $-3\sigma\geq\Delta\Delta G>-2\sigma$
for stabilizing (blue) and $\Delta\Delta G>-3\sigma$ for highly stabilizing
(navy). On the top left-hand side we show the structure of the alpha-helix
near BChl 8, which can be modified in a stabilizing way, and on the
right-hand side we highlight neighboring residues in another alpha
helix where mutations are more destabilizing. \label{fig:DDG_landscape-1}}
\end{figure*}

Free energy changes due to mutations vary depending on the location
in the protein complex monomers of \emph{P. aestuarii} and \emph{C.
tepidum. }We found that mutations at the interface of monomers are
destabilizing. We identify regions where mutations are more stabilizing
(green box) and regions where they are mostly destabilizing (black
box). In the top left panel of Fig. \ref{fig:DDG_landscape-1} we
show that the outer amino acids neighboring the 8-th BChl can be mutated
and this causes no stress / very little stress to the structure. On
the other hand, in the top right panel, we show that it is much more
destabilizing to mutate the inner alpha-helix (black box), which lies
at the interface with another monomer of the trimer. This could indicate
that, in this region, there was a larger priority to stabilize the
trimeric shape rather than the interaction with the eight BChl.

\subsection*{\textsf{Hamiltonian}}

Site couplings were computed using the point-dipole approximation
(PDA) and the transfer charge from electrostatic potential (TrEsp)
method \citep{Renger2013}. 

In the point-dipole approximation, the exciton energy transfer couplings
are computed according to Eq.~\ref{eq:pda_couplings-1}, where $\vec{d}_{10}^{A}$
denotes the transition dipole of the $0\rightarrow1$ transition for
molecule A. Here $R$ is the distance between the centers of the molecules
A and B. To compute the excitation energy transfer couplings in the
PDA we assumed an effective dipole strength of $\unit[37.1]{D^{2}}$,
an orientation of transition dipoles along the axis of the ND and
NB nitrogen atoms in the BChls and set $f=1$.

\begin{equation}
V_{10,01}^{\text{PDA}}=\frac{f}{4\pi\epsilon_{0}}\left(\frac{\vec{d}_{10}^{A}\cdot\vec{d}_{01}^{B}}{R^{3}}-3\frac{(\vec{d}_{10}^{A}\cdot R)(\vec{d}_{01}^{B}\cdot R)}{R^{5}}\right)\label{eq:pda_couplings-1}
\end{equation}

Excitation energy transfer couplings can also be computed from atomic
transition charges $q_{i}^{A}(1,0)$ and $q_{j}^{B}(0,1)$ that fit
the electrostatic potentials of transition charge densities. We calculated
atomic transition charges for all atoms in each of the BChls by fitting
the transition densities. The transition density cube files were obtained
using TD-DFT with the 6-31G{*} basis set and PBE0 functional in QChem
\citep{Shao2006}. For the fitting, we employed a method inspired
by the Merz-Kollman method \citep{Besler1990} and the CHELP-BOW method
\citep{Sigfridsson1998}. A random set of coordinates were chosen,
lying more than 2 $\AA$ and less than 8 $\mathrm{}{\AA}$ from any
atom. All points were weighted equally. We avoided all points lying
within the van der Waals radius as it has been shown that prioritizing
points outside of this radius led to charges that better reproduced
the electron density's higher multipoles \citep{Sigfridsson1998}.
A sampling frequency of 6000 points per atom was used.

We used the pseudoinverse method to solve the final system of equations
and singular value decomposition to find the pseudo inverse. This
approach avoids having to deal with poorly conditioned intermediate
matrices \citep{Francl1996}.

Excitation energy transfer couplings was then computed by summing
over the Coulomb potentials between all pairs of atomic transition
charges in the two molecules (see Eq.~\ref{eq:tresp_couplings-1})

\begin{equation}
V_{10,01}^{\text{TrEsp}}=\frac{1}{4\pi\epsilon_{0}}\sum_{i=1}^{N}\sum_{j=1}^{M}\frac{q_{i}^{A}(1,0)q_{j}^{B}(0,1)}{|\vec{r_{i}}-\vec{r_{j}}|}\label{eq:tresp_couplings-1}
\end{equation}

Excitation energy transfer couplings for all three FMO complexes calculated
in the PDA and with the TrEsp method are reported in Tab.~\ref{tab:Site-couplings-in-1}.

\begin{table*}
\centering{}\caption{Site couplings in \emph{C. tepidum}, \emph{P. aestuarii} and the ancestor
in units of $\text{cm}{}^{-1}$. Couplings were computed by using
the point-dipole approximation, with $f=1\,$ according to Eq.~\ref{eq:pda_couplings-1},
and also obtained from the transition charge using the electrostatic
potential (TrEsp) method, according to Eq.~\ref{eq:tresp_couplings-1}.\label{tab:Site-couplings-in-1}}
\begin{tabular}{ccccccc}
\toprule 
 & \multicolumn{2}{c}{\emph{C. tepidum}} & \multicolumn{2}{c}{\emph{P. aestuarii}} & \multicolumn{2}{c}{\emph{Ancestor}}\tabularnewline
\midrule 
$H_{nm}$ & Tresp {[}$\text{cm}{}^{-1}${]} & Point dipole {[}$\text{cm}{}^{-1}${]} & Tresp {[}$\text{cm}{}^{-1}${]} & Point dipole {[}$\text{cm}{}^{-1}${]} & Tresp {[}$\text{cm}{}^{-1}${]} & Point dipole {[}$\text{cm}{}^{-1}${]}\tabularnewline
\midrule
1,2 & -133.65 & -101.73 & -102.28 & -92.62 & -122.74 & -99.83\tabularnewline
1,3 & -9.53 & 7.16 & -7.89 & 7.11 & -5.53 & 4.88\tabularnewline
1,4 & -5.58 & -6.27 & 4.97 & -6.49 & -6.51 & -6.87\tabularnewline
1,5 & 7.99 & 7.65 & 8.56 & 9.17 & -9.44 & 8.54\tabularnewline
1,6 & -17.05 & -15.02 & 20.83 & -34.04 & 16.27 & -23.16\tabularnewline
1,7 & 5.31 & -9.38 & -11.33 & -15.35 & -3.05 & -8.37\tabularnewline
1,8 & 0.13 & -0.04 & 0.38 & 0.81 & 0.08 & -0.03\tabularnewline
2,3 & -39.54 & 37.13 & -39.99 & 37.48 & -37.17 & 35.63\tabularnewline
2,4 & 9.70 & 9.26 & -9.69 & 10.92 & 7.60 & 8.14\tabularnewline
2,5 & 2.06 & 2.55 & 2.11 & 1.79 & -1.80 & 1.56\tabularnewline
2,6 & 11.27 & 12.84 & -17.60 & 18.87 & -7.25 & 11.24\tabularnewline
2,7 & -3.52 & -3.66 & 3.59 & 3.60 & 2.60 & -11.89\tabularnewline
2,8 & 1.57 & 1.41 & 0.79 & 0.88 & 2.00 & 1.98\tabularnewline
3,4 & 124.56 & -108.99 & -125.38 & -129.88 & 80.99 & -81.02\tabularnewline
3,5 & -4.60 & -2.26 & 1.80 & -3.91 & -3.41 & -5.48\tabularnewline
3,6 & 13.67 & -12.17 & -11.42 & -12.26 & -12.23 & -12.45\tabularnewline
3,7 & 27.22 & 20.64 & 1.79 & 14.79 & 7.73 & -2.21\tabularnewline
3,8 & -2.98 & 2.86 & -2.89 & 2.60 & -2.75 & 2.69\tabularnewline
4,5 & -85.36 & -102.29 & 103.72 & -95.63 & 115.46 & -103.75\tabularnewline
4,6 & -26.06 & -26.00 & -25.68 & -23.24 & 27.75 & -26.56\tabularnewline
4,7 & 42.09 & -66.42 & 62.51 & -62.86 & -49.82 & -67.90\tabularnewline
4,8 & -0.86 & -2.09 & 1.90 & -2.12 & -1.91 & -2.36\tabularnewline
5,6 & 103.34 & 110.91 & -86.77 & 94.11 & 101.94 & 115.59\tabularnewline
5,7 & -2.96 & 9.09 & 18.89 & 9.54 & 4.87 & 12.58\tabularnewline
5,8 & 7.25 & 7.06 & 7.45 & 7.15 & -7.71 & 7.28\tabularnewline
6,7 & -68.02 & 51.44 & -42.72 & 56.58 & -46.75 & 60.58\tabularnewline
6,8 & -9.36 & -5.36 & 3.17 & -2.33 & 7.05 & -5.57\tabularnewline
7,8 & 9.59 & -9.79 & -10.38 & -9.50 & -10.76 & -10.21\tabularnewline
\bottomrule
\end{tabular}
\end{table*}

\begin{table}
\centering{}\caption{Average distances and standard deviations of BChl-protein distance.
The distances where computed between Mg atoms for each BChl and their
neighboring protein environment. Distances where computed for BChls
in \emph{C. tepidum}, \emph{P. aestuarii }and the ancestor. For the
protein environment we selected the $\epsilon$ nitrogen atoms of
HIS103, HIS290, HIS283 and HIS138 for sites 1, 3, 4 and 6, the $\delta$
nitrogens of HIS289 for site 7 and the peptide bond oxygens of TYR234
and TYR116 for sites 5 and 8. Residue indices refer to the \emph{C.
tepidum} sequences. Site 2 was not included as it does not bind to
the protein scaffold. Distances were computed and averaged over the
$\unit[40]{ps}$ production run.\label{tab:Average-distances-and}}
\begin{tabular}{cccc}
\toprule 
 & \multicolumn{3}{c}{Average BChl to protein distance $\left\langle d_{BChl,\,Protein}\right\rangle $}\tabularnewline
\midrule 
BChl & \emph{C. tepidum $\mathrm{[\AA]}$} & \emph{P. aestuarii} \emph{$\mathrm{[\AA]}$} & \emph{Ancestor $\mathrm{[\AA]}$}\tabularnewline
\midrule
1 & $2.81\pm0.22$ & $2.82\pm0.31$ & $2.90\pm0.26$\tabularnewline
3 & $4.31\pm0.37$ & $4.22\pm0.37$ & $3.89\pm0.26$\tabularnewline
4 & $2.73\pm0.18$ & $2.58\pm0.17$ & $2.66\pm0.22$\tabularnewline
5 & $2.47\pm0.17$ & $2.65\pm0.21$ & $2.94\pm0.29$\tabularnewline
6 & $2.79\pm0.24$ & $2.94\pm0.30$ & $2.59\pm0.18$\tabularnewline
7 & $3.82\pm0.26$ & $3.85\pm0.66$ & $3.65\pm0.30$\tabularnewline
8 & $2.23\pm0.13$ & $2.43\pm0.16$ & $2.34\pm0.16$\tabularnewline
\bottomrule
\end{tabular}
\end{table}

\subsection*{\textsf{Spectral density}}

\begin{figure*}
\begin{centering}
\includegraphics[angle=90,height=0.8\paperheight]{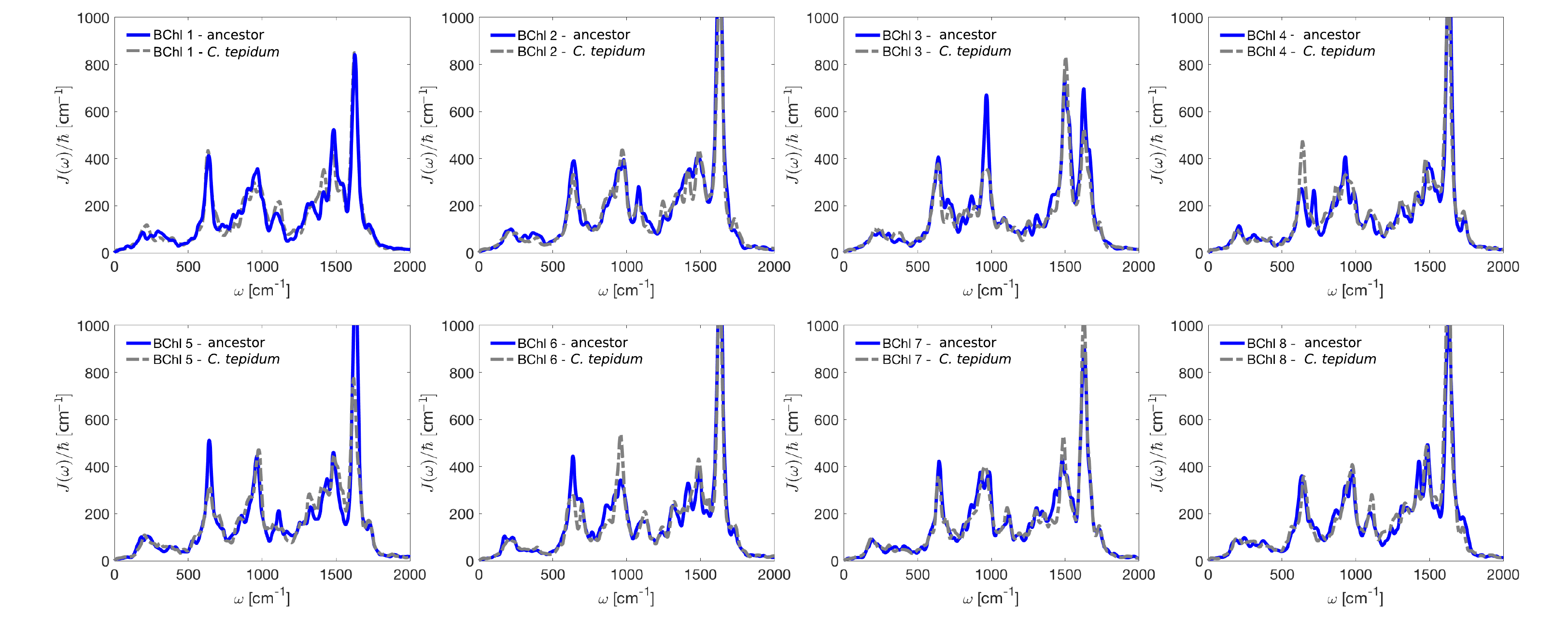}
\par\end{centering}
\caption{The spectral density, $J(\omega)$, of each BChl in the harmonic approximation
as a function of frequency. Each plot shows a comparison of the spectral
density for the \emph{C. tepidum }FMO complex to that of the ancestral
one. No large differences are observed.\label{fig:Spectral-density-in}}
\end{figure*}

\begin{figure*}
\centering{}\includegraphics[width=2\columnwidth]{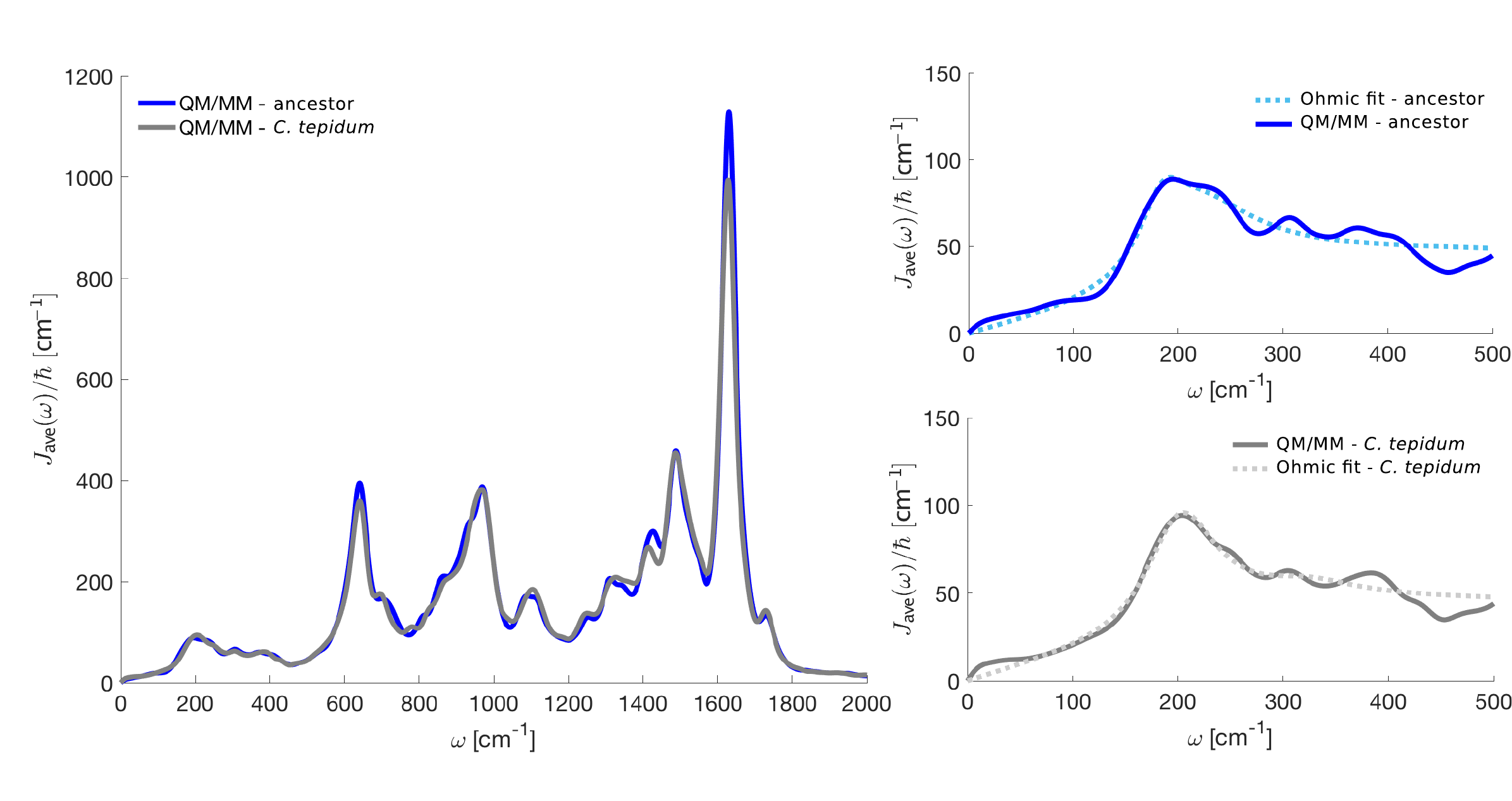}\caption{On the left hand side we show the average spectral density for the
ancestor and the average spectral density for the \emph{C. tepidum}
FMO complex. On the right hand side we show the Ohmic three-peak fit
that is taken to obtain the spectral densities used in the HEOM approach
for the dynamics. \label{fig:SD_Ohmic} The top panel shows the QM/MM
spectral density in a solid line and the Ohmic fit in a dashed line
for the ancestral FMO complex. For the bottom panel the spectral densities
are shown for \emph{C. tepidum}.}
\end{figure*}

We computed two point correlation functions using the QM/MM first
excited state energy trajectories obtained for \emph{C. tepidum} and
for the ancestor. Subsequently we fourier transformed and employed
the harmonic approximation (see Ref. \citep{Valleau2012b}) to obtain
the spectral densities $J(\omega)$. A spectral density was computed
for each BChl for both structures (see Fig. \ref{fig:Spectral-density-in}).
No large difference is seen. We also averaged the spectral densities
over all BChl's (see Fig. \ref{fig:SD_Ohmic} left panel). The two
average spectral densities are almost identical, therefore most changes
in the dynamics will originate from changes in the system hamiltonian. 

To run the dynamics of the system using the HEOM approach, we fitted
the atomistic spectral densities to a three peak Drude-Lorentz spectral
density
\begin{equation}
J(\omega)=\sum_{k=1}^{3}\left[\frac{\nu_{k}\lambda_{k}\omega}{\nu_{k}^{2}+(\omega+\Omega_{k})^{2}}+\frac{\nu_{k}\lambda_{k}\omega}{\nu_{k}^{2}+(\omega-\Omega_{k})^{2}}\right],
\end{equation}
as done, e.g. in Ref. \citep{Kreisbeck2013}. The fitted parameters
are given in Tab. \ref{tab:Drude-Lorentz-parameters-for}. The fitted
spectral densities are shown on the right hand side panels of Fig.
\ref{fig:SD_Ohmic}.

\begin{table}
\begin{centering}
\caption{Drude-Lorentz parameters used to fit the spectral density for the
ancestor and for \emph{C. tepidum.} \label{tab:Drude-Lorentz-parameters-for}}
\par\end{centering}
\centering{}%
\begin{tabular}{ccc}
\toprule 
 & \emph{C. tepidum} & \emph{Ancestor}\tabularnewline
\midrule 
$1/\nu_{1}${[}$\mathrm{fs}${]} & 66.713 & 66.713\tabularnewline
$\lambda_{1}${[}$\mathrm{cm^{-1}}${]} & 86.552 & 72.469\tabularnewline
$\Omega_{1}${[}$\mathrm{cm^{-1}}${]} & 1.0000 & 94.462\tabularnewline
\midrule 
$1/\nu_{2}${[}$\mathrm{fs}${]} & 809.03 & 506.44\tabularnewline
$\lambda_{2}${[}$\mathrm{cm^{-1}}${]} & 13.142 & 11.502\tabularnewline
$\Omega_{2}${[}$\mathrm{cm^{-1}}${]} & 199.53 & 215.99\tabularnewline
\midrule 
$1/\nu_{3}${[}$\mathrm{fs}${]} & 583.76 & 1229.0\tabularnewline
$\lambda_{3}${[}$\mathrm{cm^{-1}}${]} & 1.6977 & 5.1297\tabularnewline
$\Omega_{3}${[}$\mathrm{cm^{-1}}${]} & 320.78 & 183.72\tabularnewline
\bottomrule
\end{tabular}
\end{table}

\clearpage{}

\subsection*{\textsf{Absorption, CD and LD spectra}}

\begin{figure*}
\centering{}\includegraphics[width=1.1\columnwidth]{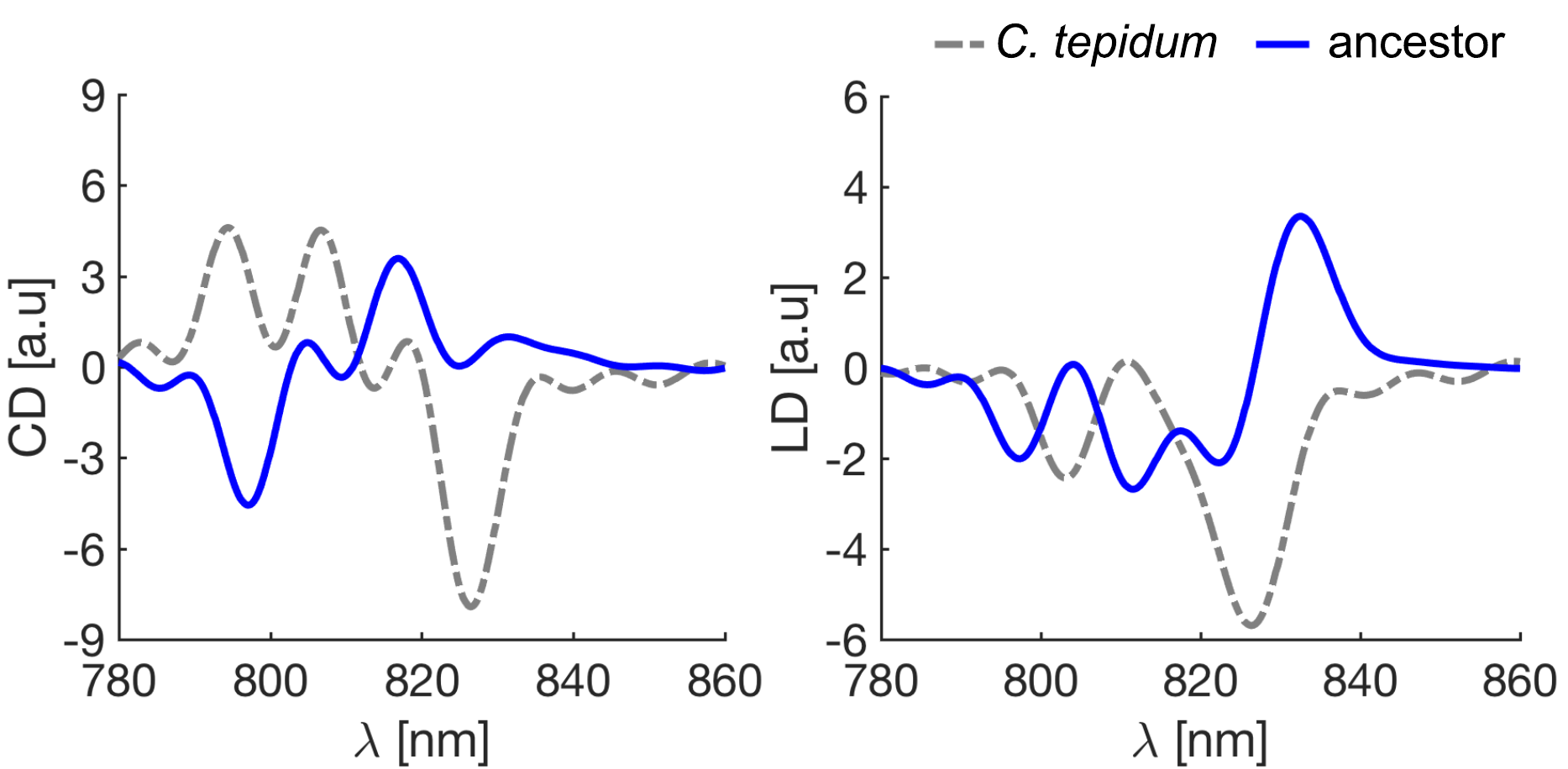}\caption{Comparison of simulated CD and LD spectra of the ancestral FMO complex
(blue line) to that of current day \emph{C. tepidum} at 300K (grey
dashed line) in arbitrary units. Spectra were computed using the Qy
transitions obtained by using QM/MM. \label{fig:Comparison-of-absorption-1-1}}
\end{figure*}

\begin{figure}
\begin{centering}
\includegraphics[width=0.9\columnwidth]{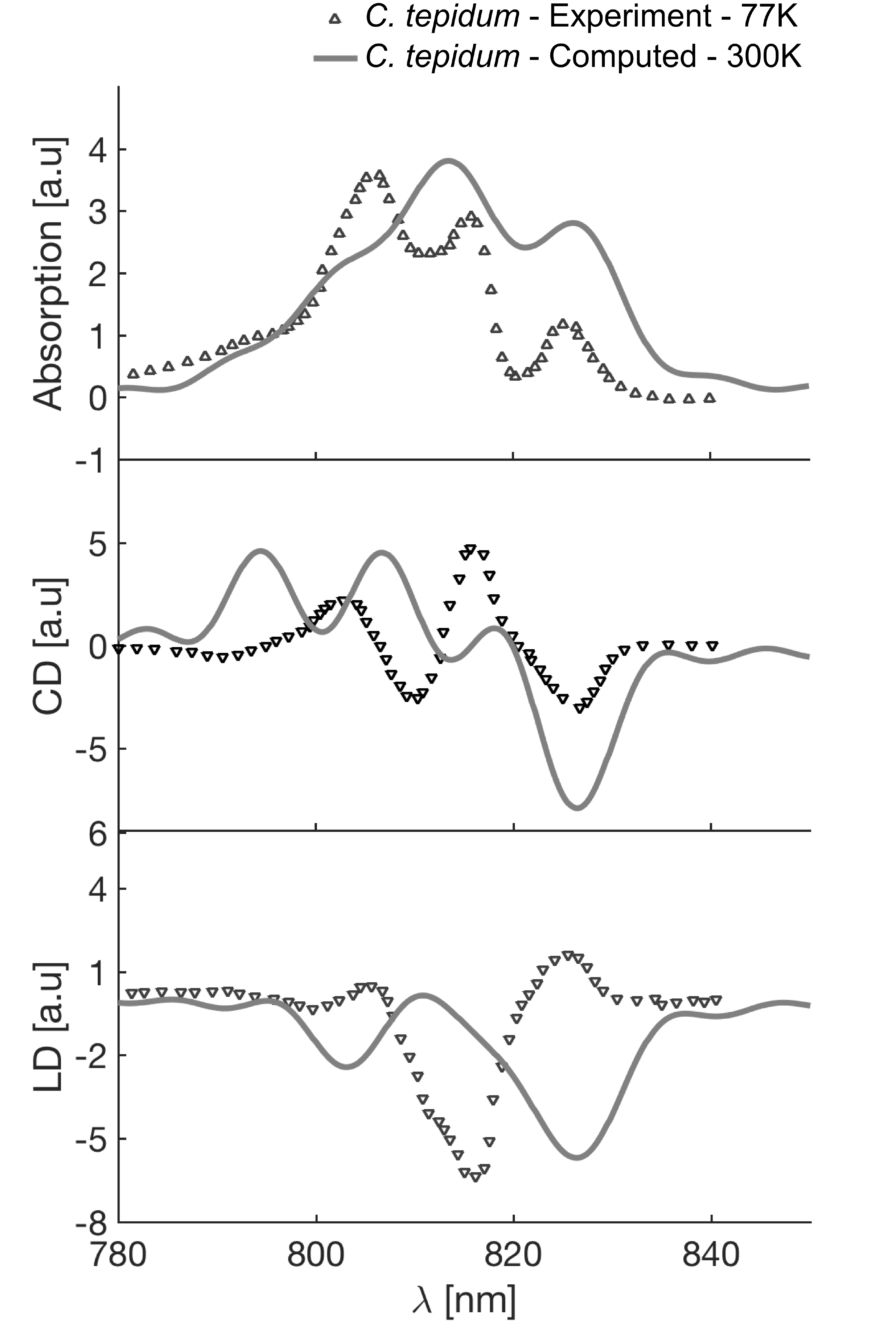}
\par\end{centering}
\caption{Comparison of our QM/MM spectra to the experimental spectra for \emph{C.
tepidum} of Ref. \citep{Vulto1998}.\label{fig:Comparison-of-our}}
\end{figure}

In Fig. \ref{fig:Comparison-of-absorption-1-1} we show the computed
CD and LD spectra for the ancestor and for \emph{C. tepidum} at 300K.
In Fig. \ref{fig:Comparison-of-our} we show the comparison of our
absorption, CD and LD spectra computed using QM/MM results at 300K
to the experimental spectra of Ref. \citep{Vulto1998}, for \emph{C.
tepidum}. We observe similar peaks in the absorption spectrum although
the 300K spectrum is much broader. The fact that the spectrum is broader
at 300K is expected because higher temperatures lead to more motion
and thus more noise in the system. We also have a similar trend for
the CD and LD spectra as that seen experimentally. 

\subsection*{\textsf{Population dynamics}}

\begin{figure}
\begin{centering}
\includegraphics[width=1\columnwidth]{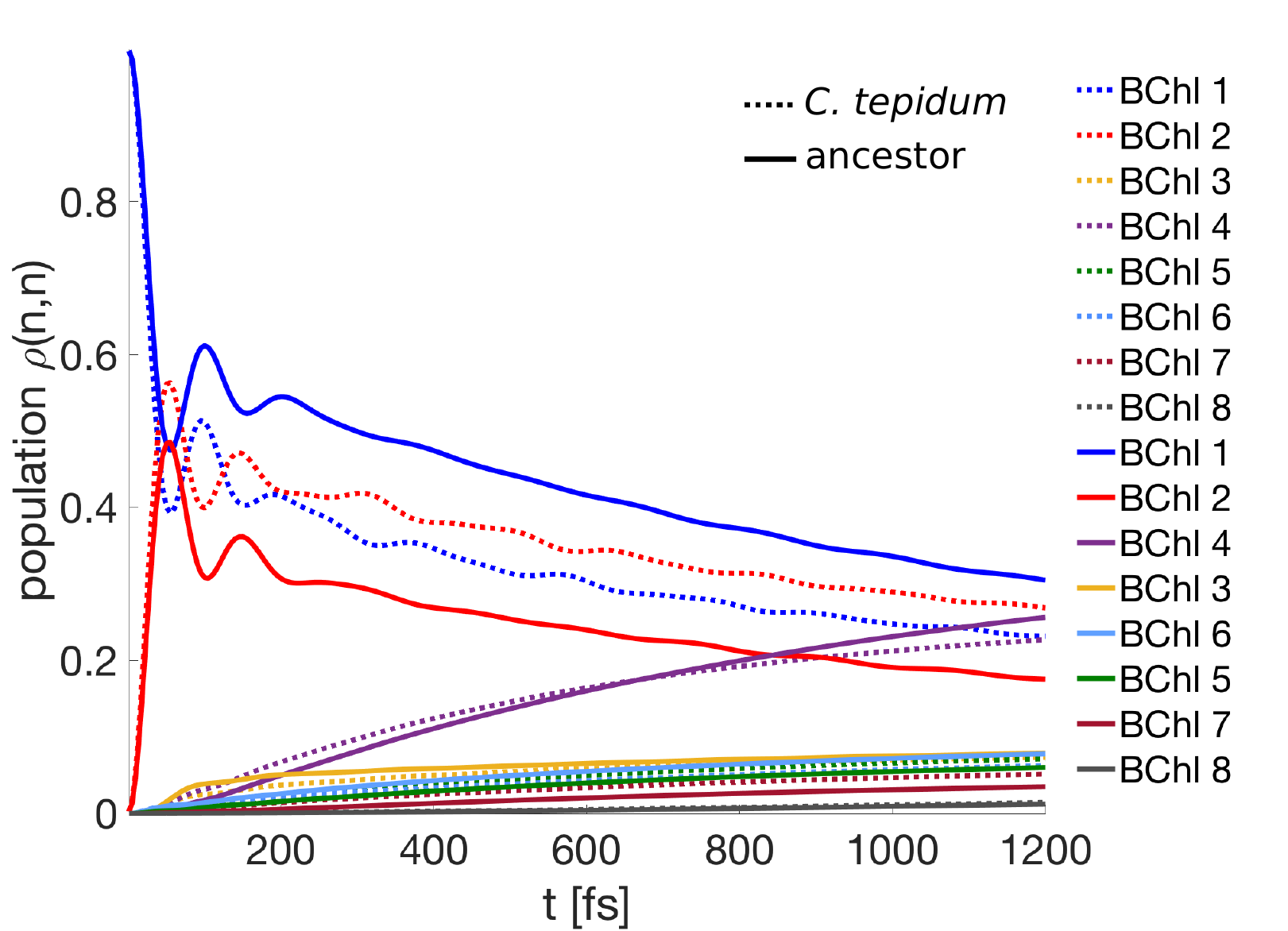}
\par\end{centering}
\caption{Excited state population of each BChl as a function of time for an
initial excited state in BChl 1 (dashed line, \emph{C. tepidum}, solid,
Ancestor) at 300K.\label{fig:Panel-a)-Population-1}}
\end{figure}

\begin{figure}
\begin{centering}
\includegraphics[width=1\columnwidth]{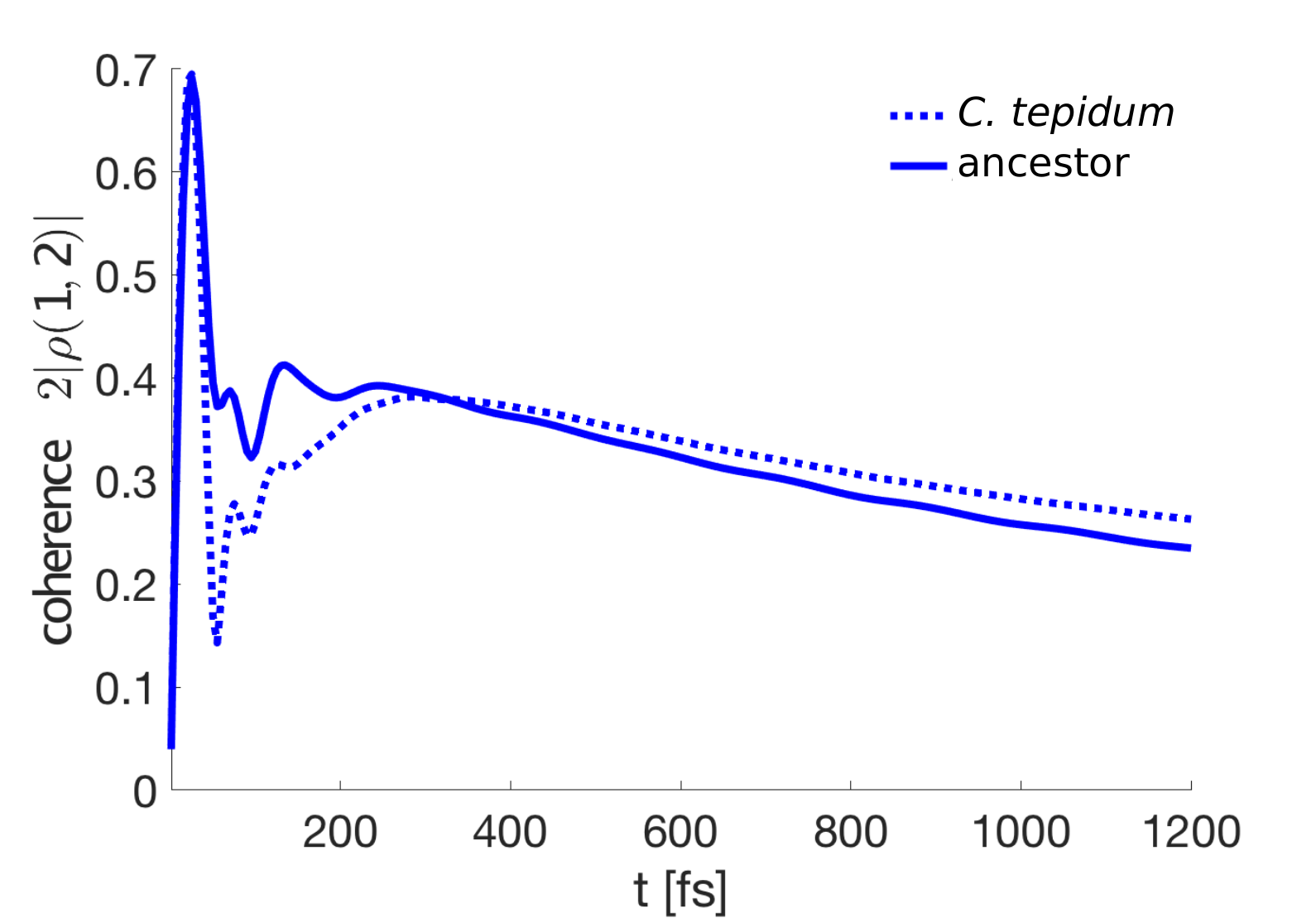}
\par\end{centering}
\caption{Coherence between the excited states of BChls 1 and 2 as a function
of time. These results were obtained by propagating the density matrix
of the system using HEOM at 300K and with BChl 1 as the initial state.
\label{fig:Coherence-between-excited}}
\end{figure}

Using the HEOM approach \citep{Tanimura1989,Tanimura2006,Ishizaki2009a}
we computed the time evolution of the density matrix of our system
of 8 coupled BChl molecules using \textit{QMaster} \citep{Kreisbeck2014,KreisbeckCode}.
The spectral densities were taken to be three peak fits of the atomistic
spectral density, as described previously. The resulting population
for the ancestor and \emph{C. tepidum} in shown in Fig. \ref{fig:Panel-a)-Population-1},
we see that site 1 and 2 are less strongly coupled for the ancestor
than they are for \emph{C. tepidum}. One can see this also from the
Hamiltonians as given in Tab. \ref{tab:Site-couplings-in-1}. The
coherence between site 1 and 2 is shown in Fig. \ref{fig:Coherence-between-excited}.
Here we do not see a large difference between the two species and
this further confirmes the fact that quantum coherence does not appear
to have been selected for.

\end{document}